\documentclass[sigconf]{acmart}

\usepackage{subcaption}

\AtBeginDocument{%
  }

\setcopyright{acmlicensed}
\copyrightyear{2024}
\acmYear{2024}
\acmDOI{XXXXXXX.XXXXXXX}

\acmConference[CHI '25]{Proceedings of the 2025 CHI Conference on Human Factors in Computing Systems}{April 26--May 01,
  2025}{Yokohama, Japan}
\acmISBN{978-1-4503-XXXX-X/18/06}

\usepackage{docmute}
\usepackage{listings}
\DeclareFixedFont{\ttb}{T1}{txtt}{bx}{n}{12} 
\DeclareFixedFont{\ttm}{T1}{txtt}{m}{n}{12}  

\usepackage{color}
\definecolor{deepblue}{rgb}{0,0,0.5}
\definecolor{deepred}{rgb}{0.6,0,0}
\definecolor{deepgreen}{rgb}{0,0.5,0}

\newcommand\pythonstyle{\lstset{
language=Python,
basicstyle=\ttfamily\footnotesize,
morekeywords={self},              
keywordstyle=\ttb\footnotesize\color{deepblue},
emph={MyClass,__init__},          
emphstyle=\ttb\footnotesize\color{deepred},    
stringstyle=\color{deepgreen},
frame=tb,                         
showstringspaces=false,
breaklines=true,
}}

\lstnewenvironment{python}[1][]
{
\pythonstyle
\lstset{#1}
}
{}

\newcommand\pythonexternal[2][]{{
\pythonstyle
\lstinputlisting[#1]{#2}}}

\newcommand\pythoninline[1]{{\pythonstyle\lstinline!#1!}}

\definecolor{codegreen}{rgb}{0,0.6,0}
\definecolor{codegray}{rgb}{0.5,0.5,0.5}
\definecolor{codepurple}{rgb}{0.58,0,0.82}
\definecolor{backcolour}{rgb}{0.95,0.95,0.92}
\definecolor{backcolour_prompts}{rgb}{0.92,0.91,0.62}

\lstdefinestyle{mystyle}{
    backgroundcolor=\color{backcolour},   
    commentstyle=\color{codegreen},
    keywordstyle=\color{magenta},
    numberstyle=\tiny\color{codegray},
    stringstyle=\color{codepurple},
    basicstyle=\ttfamily\footnotesize,
    breakatwhitespace=false,         
    breaklines=true,                 
    captionpos=b,                    
    keepspaces=true,                 
    numbersep=5pt,                  
    showspaces=false,                
    showstringspaces=false,
    showtabs=false,                  
    tabsize=2
}

\lstdefinestyle{prompts}{
    backgroundcolor=\color{backcolour_prompts},   
    commentstyle=\color{codegreen},
    keywordstyle=\color{magenta},
    numberstyle=\tiny\color{codegray},
    stringstyle=\color{codepurple},
    basicstyle=\ttfamily\footnotesize,
    breakatwhitespace=false,         
    breaklines=true,                 
    captionpos=b,                    
    keepspaces=true,                 
    numbersep=5pt,                  
    showspaces=false,                
    showstringspaces=false,
    showtabs=false,                  
    tabsize=2
}

\begin{document}

\title{DreamGarden: A Designer Assistant for Growing Games from a Single Prompt}

\author{Sam Earle}
\email{sam.earle@nyu.edu}
\orcid{https://orcid.org/1234-5678-9012}
\affiliation{%
  \institution{New York University \& Microsoft}
  \country{USA}
}

\author{Samyak Parajuli}
\email{samyak.parajuli@utexas.edu}
\affiliation{%
  \institution{UT Austin}
  \country{USA}
}

\author{Andrzej Banburski-Fahey}
\email{abanburski@microsoft.com}
\affiliation{%
  \institution{Microsoft}
  \country{USA}
}


\begin{abstract}
Coding assistants are increasingly leveraged in game design, both generating code and making high-level plans. To what degree can these tools align with developer workflows, and what new modes of human-computer interaction can emerge from their use? We present DreamGarden, an AI system capable of assisting with the development of diverse game environments in Unreal Engine. At the core of our method is an LLM-driven planner, capable of breaking down a single, high-level prompt---a dream, memory, or imagined scenario provided by a human user---into a hierarchical action plan, which is then distributed across specialized submodules facilitating concrete implementation. This system is presented to the user as a garden of plans and actions, both growing independently and responding to user intervention via seed prompts, pruning, and feedback. Through a user study, we explore design implications of this system, charting courses for future work in semi-autonomous assistants and open-ended simulation design.

\end{abstract}

\ccsdesc[500]{Human-centered computing~Human computer interaction (HCI)}
\ccsdesc[300]{Applied computing~Media arts}
\ccsdesc[100]{Computing methodologies~Multi-agent planning}

\keywords{Game design assistants, 3D asset generation, large language models, visual feedback}

\begin{teaserfigure}
  \includegraphics[width=\textwidth]{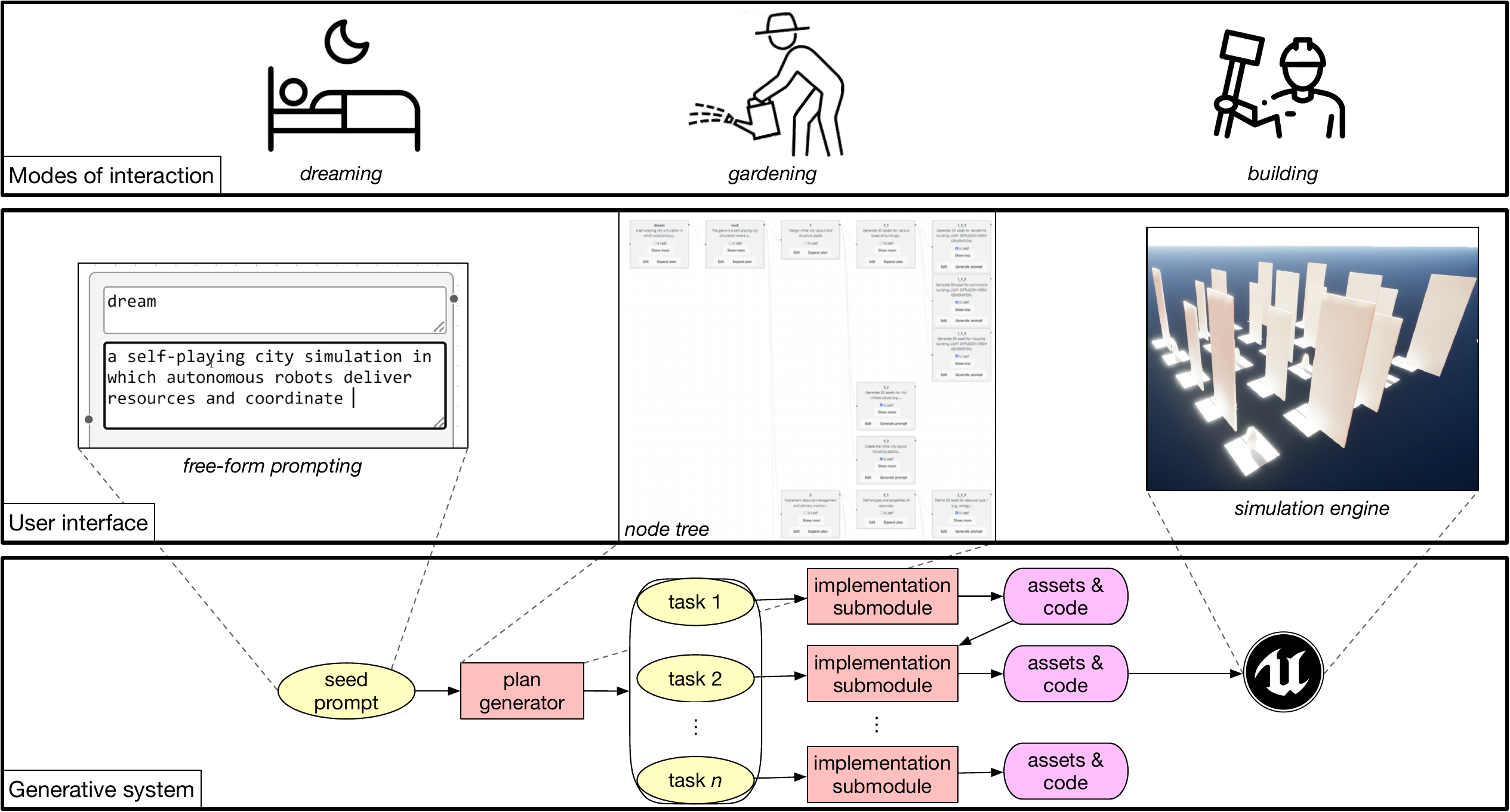}
\caption{\textit{DreamGarden} is a game design assistant facilitating diverse modes of user interaction. Users enter open-ended ideas into a planning module which recursively refines them into actionable design plans---a garden to be pruned or expanded. Implementation submodules carry out the resultant sequence of tasks, iterating on their work via autonomous visual evaluations or by responding to direct user feedback and code edits.}
\label{fig:frontfig}
\end{teaserfigure}


\received{20 February 2007}
\received[revised]{12 March 2009}
\received[accepted]{5 June 2009}

\maketitle


\section{Introduction}


A recent boom in generative AI has led to the widespread adoption of assistive AI tools by human creators, both casual and professional, in generating text, code, images and 3D assets.
The AI models underpinning these tools are generally large neural networks, trained on corpora of human-generated content scraped from the internet.
Large language models (LLMs) are perhaps most ubiquitous, existing most commonly in the form of chatbots, predictive text models which have been fine-tuned via specialized datasets and human feedback to be generally ``helpful'' and excel in following user requests \cite{openai2024chatgpt, Touvron2023Llama2O, Jiang2023Mistral7}.
In the visual domain, diffusion models have brought on similarly impressive capabilities, allowing casual users to create visually convincing results from a single prompt \cite{rombach2022high, Nichol2021GLIDETP, Ramesh2022HierarchicalTI}.

Generally, these assistants turn individual user prompts into output wholesale, and users have little means of intervening in the process beyond asking the model to try again with adjusted prompts.
But this prompt-to-output paradigm becomes impractical as creative applications become more complex.
Designers of games or simulated environments (e.g. for film, architecture, or VR training) are unlikely to forego their domain expertise---and in particular their mastery of symbolic languages for describing artifacts---in favor of lengthy iterated prompting dialogues with fallible black box generators.
Nonetheless, the capabilities and knowledge of large models offer massive potential for assisting designers in such domains, and so a major question in unlocking these capabilities is what interaction paradigm befits such scenarios.

Another obstacle toward leveraging assistants in complex domains is the sheer complexity of the tasks themselves.
As the number of possibilities for error increases exponentially, the amount of back-and-forth that would be required between a user and the model in a straight-ahead prompting paradigm would result in intolerable user fatigue. It becomes necessary for the system to iterate autonomously to some degree to meet the user's desiderata and work within their constraints, interfacing with design tools to evaluate and refine its own work without the need for constant user oversight.

\begin{figure*}
\centering
\includegraphics[width=1.0\linewidth]{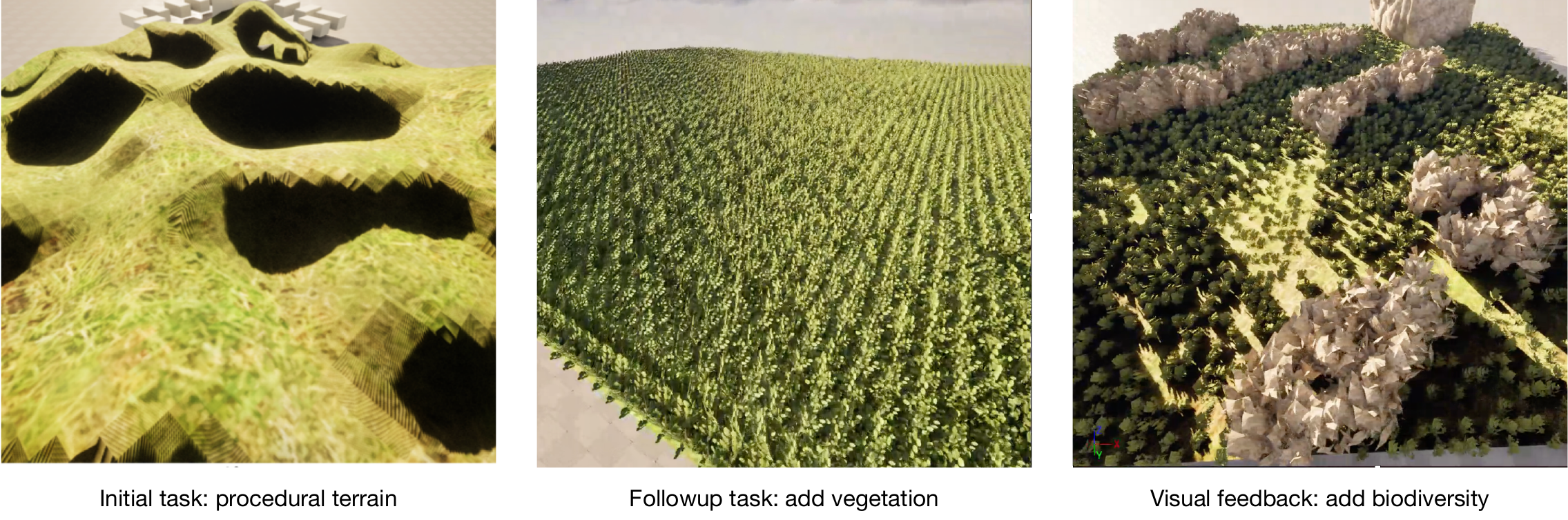}
\caption{Growth of a scene in DreamGarden. The initial task prompts the procedural mesh generation submodule to write code using Perlin noise to produce hilly terrain (left). A followup task requests foliage to be added to this terrain (middle) which is rendered more diverse after visual feedback (right).}
\label{fig:terrain_iteration}
\end{figure*}

The automated game design assistant we present in this paper, \textit{DreamGarden}, represents a possible answer to these questions. The core idea is to recursively prompt an LLM to devise a hierarchical plan of action, setting its own sub-goals which can be iterated on autonomously while interfacing with the design tool in question (in our case, a game engine). This tackles the problem of complexity by subdividing tasks and allowing for autonomous iteration without requiring the user to act as a go-between for model output and the design tool. Meanwhile, the system's plan and iteration attempts are exposed to users via a graphical interface, allowing them to intervene at a more granular level beyond the initial high-level prompt, but still at a higher level of abstraction than directly editing generated artifacts (though this is also a possibility within our system).

As a running illustrative example, we will consider a case in which a user seeks to generate ``a simulation of cows and other animals grazing in a field around a farm, where a spaceship flies above the animals, intermittently abducting them''. The system first rephrases this loose, natural language prompt in more concrete terms given the tools it has at hand. Namely, the system may ultimately generate 3D assets using a  text-to-2D-to-3D diffusion based pipeline, download existing 3D assets from an extensive online database by searching for asset thumbnails most aligned with a generated text description via a contrastive text-image loss, write arbitrary Unreal Engine~\cite{unrealengine} code (in the form of C++ ``Actor'' classes that represent any game entities, from an individual NPC to a spawner, an abduction beam, a terrain object, etc.) which may instantiate these assets, and place these actors in some initial layout in a scene. 

With these implementation tools in mind, the language model breaks down the task into a hierarchical plan which is represented as a tree stemming from the user's root prompt. At the leaves of this tree are the implementation tasks themselves, which are then executed by individual implementation submodules: e.g. assets will be downloaded until a valid close match is found, and code will be re-written until it compiles and produces visually satisfactory intermediate results. Once each implementation task is completed, the cumulative output of the system---the generated assets, code, and scene layout---is transferred to subsequent implementation submodules by injecting them into their prompts where appropriate. In our running example, the system writes code to procedurally generate some grassy naturalistic terrain, then generates a spaceship mesh via diffusion, then writes additional code to have this spaceship fly above the terrain semi-randomly, and so on.

At any point during the system's operation, the user may choose to prune or expand nodes within the plan tree, either simplifying and condensing subtasks or breaking them out into further child tasks, respectively. The user may also provide feedback on any of the system's generated intermediary output, which is then taken into account in subsequent implementation steps to help  guide the generation process. The user may also intervene directly, for example by editing code or providing alternative or additional assets, which changes will similarly be fed into the system at subsequent steps in the generation process. In this work, we focus on refining the system's autonomous capabilities and in exploring user experience of these modes of possible interaction. Meanwhile, the general idea of a node-based interface of plan steps and implementation attempts opens up myriad possibilities for future work wherein users could freely edit, recombine and reorganize plan and action nodes to ideate and iterate on arbitrary games and simulations.

In sum, we introduce \textit{DreamGarden}, an AI-driven assistant for game design, which works semi-autonomously toward a high-level goal by growing a ``garden'' of plans and actions, where intermediate output is human-interpretable and editable.
In designing this system, we pursue a sliding scale between autonomy and reactivity. We target the early ideation and prototyping phases of game design, with the goal of enabling rapid prototyping of playable game snippets. \textit{DreamGarden} elaborates on an initial game design plan, expanding it into a sequence of concrete implementation steps which are then automatically delegated to various submodules. \textit{DreamGarden} is a first such attempt at an automated game prototyping tool to our knowledge.

To validate the system and explore its potential as a new means of human-computer interaction, we conduct a study with the following research questions in mind:\
\begin{itemize}
    \item[(Q1)] Is the system capable of transforming open-ended, potentially dream-like natural language prompts into functioning 3D simulation environments?
    \item[(Q2)] Is the system's hierarchical and iterative process intuitive and accessible to users?
    \item[(Q3)] Does DreamGarden's user interface provide ample means for users to intervene in this process, at various points in time and at various levels of abstraction?
\end{itemize}

\begin{figure*}
\includegraphics[width=1.0\textwidth]{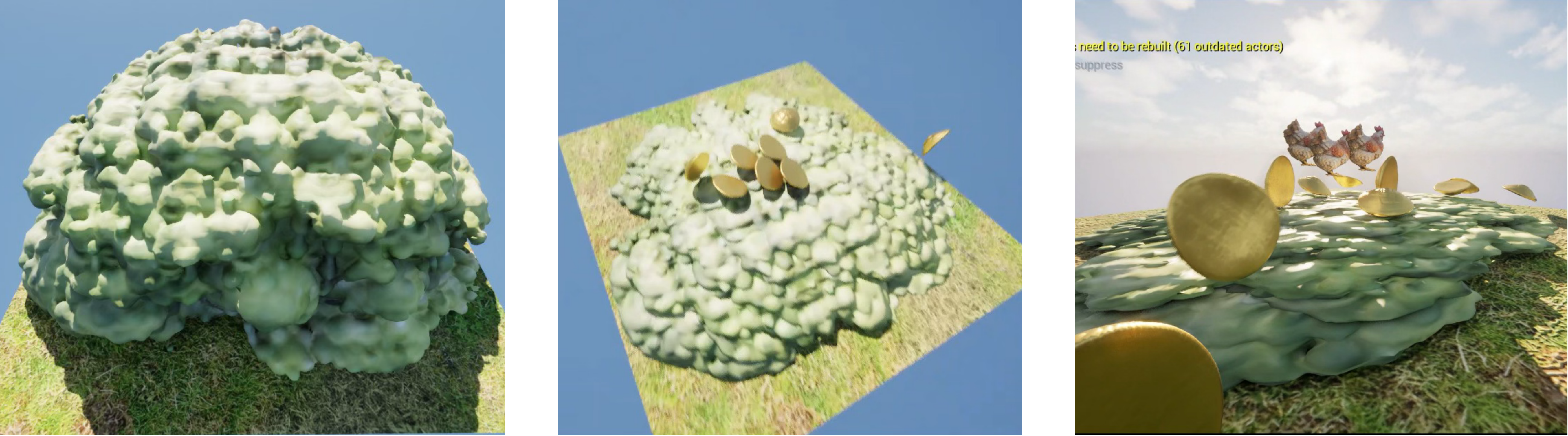} 
\caption{Example of autonomous scene-generation on DreamGarden. Here, the system iterates on a scene involving chickens laying golden eggs by writing and extending C++ Unreal Actor classes, beginning with some initial terrain and foliage, then adding dynamically-spawning and physics-enabled golden eggs, and finally chickens, using meshes that it has generated/downloaded at prior iterations.}
\end{figure*}


 

\section{Background: procedural content generation and open-ended learning}

In game design, procedural content generation (PCG) has long been leveraged to various ends; in turn 
affording designers a new means of expression and a higher level of abstraction when designing content, accelerating the design process, and producing novel and surprising artifacts leading to increased replayability from the perspective of the player \cite{kreminski2023generator}.
Such approaches move designers away from generating game artifacts such as levels ``brick by brick'' and instead allow them to express slightly more general and abstract constraints on their domain. For example, constraint-solving algorithms~\cite{cooper2022sturgeon} or wave function collapse~\cite{Gumin_Wave_Function_Collapse_2016}  require designers to predefine a set of game components and/or rules dictating their possible spatial relationships in a level in order to guarantee the functionality/playability of the generated content. 

More recently, a large body of research has explored PCG via Machine Learning (PCGML~\cite{summerville2018procedural}), wherein supervised learning is used to train a model on datasets of human-generated content, which effectively learns to generate novel but plausible artifacts at runtime.
In many cases, a spatial artifact---such as a tile-based level---is likened to an image, and approaches from computer vision are adapted to generate levels~\cite{awiszus2020toad, merino2023five}.
Unlike more traditional methods which may involve costly backtracking to search for satisfactory content, or alternative methods like evolution, which require a massive number of samples to produce individual artifacts~\cite{shaker2012evolving}, learned generators have the advantage of being fast at runtime. 
However, game content is generally beholden to stricter constraints than natural images.
Moreover, these constraints may vary widely between games, with datasets of viable content for a given game being generally far more limited than corpora of natural images.
One major obstacle to PCGML is thus the need for learning and enforcing functional constraints.
\cite{zhang2020video}, for example, address this when applying the Generative Adversarial Network (GAN) popularized in computer vision to the generation of tile-based dungeons, by augmenting the GAN with a Mixed Integer Programming-based constraint solver in order to find the functionally viable levels nearest the aesthetically plausible output learned by the generator.

A desire for generators capable of rapidly generating a diversity of plausible content at runtime while obeying functional constraints is what motivates PCG via Reinforcement Learning (PCGRL~\cite{khalifa2020pcgrl}), in which, instead of training on human data, designers specify functional or aesthetic desiderata via computable heuristic functions. Generators are then trained via Reinforcement Learning (RL) to maximize reward functions corresponding to weighted linear combinations of these heuristics while iteratively making local edits to the level (effectively transforming the task of game level design into a game in itself). PCGRL has been extended to train controllable generators, where a designer can tune the target values of relevant metrics at run-time in order to elicit controllably diverse content from generators~\cite{earle2021learning}, and additionally applied to rudimentary 3D domains~\cite{jiang2022learning}.

While PCGML approaches relying only on human data are liable to produce functionally invalid artifacts, PCGRL, which explicitly optimizes for functional constraints, has no way of capturing the ``human'' quality of level design. Consider, for example, the application of PCGRL-like methods in Open-Ended Learning ~\cite{dennis2020emergent, bontrager2021learning}. In such scenarios, PCGRL is used to generate game levels to provide optimal challenges for learning player agents. While this line of work seeks to address the widespread issue of overfitting in RL by widening the distribution of environments online and adaptively, there is no guarantee the environments providing optimal challenges for agents will be of interest to humans, or whether they will be diverse enough 
to engender RL agents capable of generalizing to the set of unseen real-world environments in which the agents may ultimately be deployed.

Several possible approaches present themselves to address the issue of alignment in Open-Ended Learning and in learned level generators more generally.
Notably, OMNI-EPIC~\cite{faldor2024omni} leverages pre-trained language models to assist in writing novel 3D RL environments providing challenges for embodied agents, enabling training agents in a wide array of rudimentary but human interpretable 3D games in a simple physics-based engine.
Without the ability of large models to understand and remix the typical conventions of 3D game design, generating this variety of environments would have either require a heavily constrained representation of the environment, limiting environmental diversity;
or a fine-grained representation space (e.g. evolving arbitrary strings of code), admitting too many inadmissible environments to make naive search feasible.

\textit{DreamGarden} is adjacent to this line of work in aligning open-ended learning to notions of human interest via guidance by large pre-trained models.
While the environments it produces in Unreal Engine are not RL environments per se, the system demonstrates a pipeline for automatically generating a rich space of game environments, diverse in terms of their mechanics, which could ultimately serve as training grounds for embodied agents.
Moreover, while OMNI-EPIC's generated worlds are human-interpretable, \textit{DreamGarden}'s are particularly semantically rich, with a focus on narrative, real- (or dream-)world scenarios, involving more visually dense and representational entities (contrasted against the basic primitives of open-ended frameworks like OMNI-EPIC or X-LAND~\cite{team2023human}).

The most boundary-pushing open-ended learning will likely not rely on guidance of  frozen human knowledge distilled into the weights of a large model, because true open-endedness should move beyond the purview of this pre-existing knowledge.
A complimentary approach is to involve humans in this guidance process, who are capable of on-the-fly adaptation to alien worlds.
In relation to Open-Ended Learning, then, \textit{DreamGarden} represents an exploration of the shape such oversight might take, involving intervention at varying levels of abstraction.
Several insightful examples of such human-in-the-loop, semi-autonomous environment generation loops already exist in the form of automated game design assistants, though they are generally restricted to families of relatively simple, tile-based games.


\textit{DreamGarden} can be seen as new approach to PCG, a kind of meta-PCG in which LLMs are prompted to define their own PCG workflows.
For now, that means generating a sequence of mesh-generation, mechanic generation and object-layout tasks.
It could be extended with tools for producing L-systems (e.g. for flora), cellular automata (for room layouts) and even training ML or RL generators with datasets or heuristics.

We may take a relatively conservative view of LLMs as capable of regurgitating and sometimes interpolating human artifacts.
As a designer tool, \textit{DreamGarden} asks how we can get a more intuitive sense of what this possibility space feels like.\footnote{One issue is that pre-training may have a tendency toward collapsing to the mean and learning a space of viscerally derivative-feeling ``AI slop''. It also comes with potential intellectual property issues. We consider these pitfalls in Section \ref{sec:ethics}.}.

\section{Related work}

\subsection{Artificial life and generative gardens}


\textit{DreamGarden} makes a designer tool (or perhaps a design game) out of a self-perpetuating garden of design ideas and content.
We take inspiration from the field of Artificial Life~\cite{aguilar2014past}, which involves the design of simulated environments in which meaningfully complex forms can emerge automatically.
Dave Ackley's Moveable Feast Machine (MFM), for example, distributes a computer program across a spacial plane involving cellular automata-like actors~\cite{266903}.
With forms like a self-healing wire, computation (though less efficient) becomes robust in biologically plausible ways.
Lenia~\cite{chan2018lenia}, a continuous extension of John Conway's cellular automaton Game of Life~\cite{izhikevich2015game} has been combined with open-ended search algorithms such as Quality Diversity search~\cite{pugh2016quality} for the automatic discovery of novel and interesting forms~\cite{faldor2024toward}.

Such systems can be likened to simplistic digital petri dishes or miniature ecosystems
When augmented with a user-facing aspect, they might be thought of as gardens.
These gardens can be communal: \cite{charity2022aesthetic} develop an online Twitterbot which seeks to autonomously generate aesthetically pleasing gridworld game maps by soliciting human feedback via automated polls.
Users are also able to design their own levels on the platform, which are pitted against output from a small generative model that learns to imitate highly-rated levels.
PicBreeder~\cite{secretan2011picbreeder} has users interactively evolve Compositional Pattern Producing Networks~\cite{stanley2007compositional}, an idiosyncratic representation of infinite-resolution images, often discovering novel forms without having set explicit objectives in advance.
Though DreamGarden is framed as a system for translating a prompt to a corresponding end product, the plans and artifacts it sometimes generates (e.g. to generate 2,000 blades of grass individually, or the expressionistic and self-colliding meshes resulting from initial attempts to procedurally generate standard objects) could potentially be explored as art objects on their own merits, with the system's intermediary outputs repurposed for divergent, open-ended search~\cite{stanley2015greatness}. 


\subsection{Generative design assistants}

\textit{DreamGarden} belongs to a rapidly growing body of work studying the application of large models as design assistants.
\cite{ling2024sketchar} develop a web-based tool which uses text-to-image models to help users prototype their design concepts, rapidly translating between text and images with an eye toward bridging the gap that can emerge between artists and designers communicating across these modalities. LLMs have been used to power an evolutionary engine for collaborative game design in \cite{lanzi2023chatgpt}. 
Such tools are contrasted against LLM-free approaches that rely on heavier constraints on content.
Sentient Sketchbook~\cite{liapis2013sentient} uses novelty search to provide alternatives to human designs, and automates playability checks; while~\cite{10.1145/1520340.1520713} develop an authoring tool for machinima, automating aspects like character control and shot framing to help novice creators readily apply conventional cinematic forms.

LLMR \cite{DeLaTorre2023LLMRRP} also comprises a multi-agent LLM-driven systems capable of translating input prompts into largely unconstrained 3D environments.
\textit{DreamGarden} is like a slower, deeper cousin of LLMR, offering less instant immersion and more branching possibility.
LLMR is a tool for casual creators~\cite{compton2015casual} enabling real-time scene and environment generation in VR for immediate use.
It thus limits itself to rapid feedback, rendering impractical the recursive prompting we employ to produce hierarchical plans, which in turn generally entail long sequences of implementation steps, each with a generous number of attempts permitted to fix bugs and refine output.
With no plan and action tree to expose to users, and making no assumptions of users' coding proficiency, LLMR involves prompt input alone.
Similarly, whereas in LLMR, interaction with the system is designed to be seamless, with intermediary prompts and code largely concealed from the user, our user interface is designed to expose this process in an intuitive way, allowing ample opportunities for designer intervention.


\begin{figure*}
    \centering
    \includegraphics[width=1.0\linewidth]{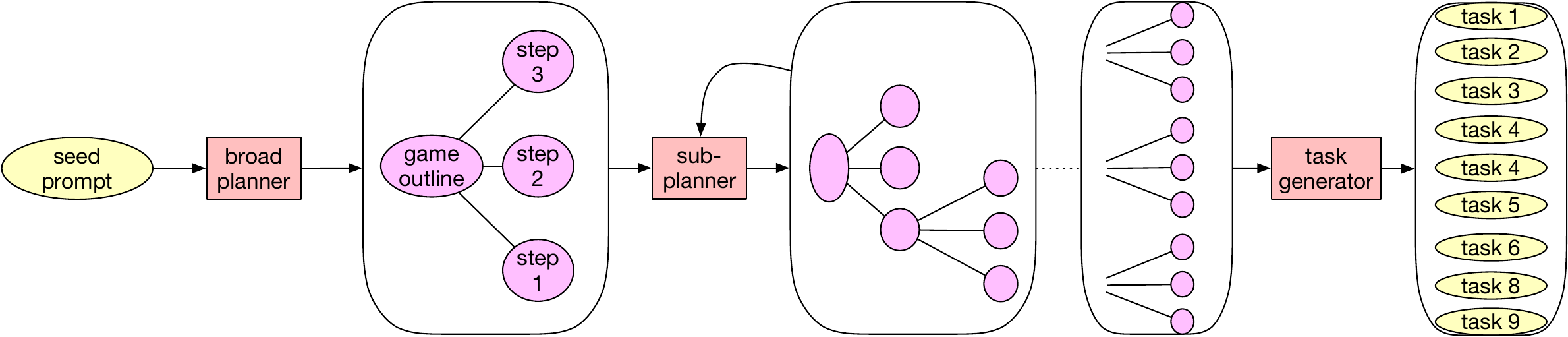}
    \caption{The planning module converts the user's open-ended seed prompt into a broad plan for designing a simulation in Unreal Engine, then recursively breaks this plan down into more fine-grained steps, eventually terminating in leaf nodes which are then converted to concrete implementation tasks for the available implementation submodules.}
    \label{fig:plan_tree}
\end{figure*}

\subsection{World models}

In RL, where game-playing agents are trained via a reward corresponding to in-game score or success, world models---generative models of game mechanics and levels~\cite{ha2018world}---have been developed to alleviate the bottleneck of running expensive game simulations during agent training, often leading to highly performant game-playing agents~\cite{hafner2020mastering}.
Generally, these models predict the next frame of the simulation given the previous frame(s) and current agent action.
Similar models have also been explored for their purely generative capability, with Genie \cite{bruce2024genie} training a model on a large internet corpus of videos of gameplay of diverse games to predict next frames and infer player actions, then prompting it at runtime to generate next frames of imagined games given out-of-distribution input such as human-drawn sketches. \cite{valevski2024diffusionmodelsrealtimegame} showed that diffusion models can serve as game engines, with a demonstration of stable simulation of the classic game Doom on a single TPU, or Counter Strike: GO on a single GPU \cite{alonso2024diffusion}, though these  approaches can be ran for relatively short times before losing coherence.

A design workflow involving such models, then, limits designer intervention to the provision of the initial prompt: the model does not manipulate any symbolic representation of the game world, and designers cannot tweak its output with any level of detail once the generation process is complete, nor do the models make use of existing methods for rendering assets or encoding game logic which might make their output more reliable and their generative task more tractable.

The system developed in this paper, by contrast, develops an interface between generative models and Unreal Engine~\cite{unrealengine}, a robust editor for game levels, assets and logic, popular both with indie developers and AAA studios.

\subsection{Automatic game design}

Prior work has explored the use of language models to generate games from scratch. Word2World~\cite{nasir2024word2world} generates narratives from scratch and subsequently uses this narrative to specify a set of characters, a tileset, and a sequence of grid-based level layouts.
Individual tiles are then generated using the text-to-image model Stable Diffusion~\cite{rombach2022high}.
Here, game mechanics and objectives are essentially fixed in advance, and limited to player movement, the impassibility of certain tile types, and achievement of certain objectives by reaching special tiles.
By contrast, DreamGarden is unrestricted in game mechanics---in terms of interactions between abstract Actor classes written in C++ ---which can be determined from scratch by our code generation submodule.

The procedural generation of entire games has been an area of research prior to the rise of modern deep learning,
with the first experiments in Automatic Game Design demonstrated in \cite{10.1007/978-3-540-74782-6_54} and \cite{5035629}.  The prospect of learning game design directly from games was proposed in \cite{osborn2017automated}. Recombining learned features for game design was considered in \cite{guzdial2018automated}, while in \cite{9231927} authors have shown how blending of different games can be achieved using ML techniques.
For overviews of the field, see \cite{10.1145/2422956.2422957, yannakakis2018artificial}.

The authors of \cite{cook2016angelina} raised pressing questions around the valuation of creativity and the participation of  autonomous agents in game development communities.
While our system could be let loose as an autonomous agent, this angle is at present somewhat less viable because of the scope of our ambition in generating fully open-ended 3D simulations via arbitrary code.
We also take for granted that---no matter how much the frontier of automatic game generation is pushed forward by the orchestration of increasingly sophisticated specialized models---the components of such a system will inevitably be repurposed as designer-facing tools, whether by experimental AI artists or seasoned developers.
In this paper we therefore focus mainly on the possible modes of interaction afforded within such systems.

Rosebud AI \cite{rosebud} serves as a notable startup in this domain, providing an AI-powered platform for creating browser-based games in JavaScript, along with a collaborative community where game developers can share and build upon each other's creations. Our method offers a similar natural language-to-code tool but focuses on structured, plan-to-action workflow. This workflow guides users through the necessary sub-tasks to fully realize their creative vision. Furthermore, while Rosebud AI focuses on 2D browser games, we specialize in the 3D domain allowing us to support more complex asset creation and immersive environments.

\subsection{Large models for code, asset and environment generation}

Alongside the nascent push to leverage large models for automatic game generation, LLMs have also been applied to game design sub-tasks, such as the generation of levels in Sokoban~\cite{todd2023level} and Super Mario Bros.~\cite{sudhakaran2024mariogpt}.
Meanwhile, large text-to-image models have been leveraged for the generation of structures and environments in MineCraft~\cite{earle2024dreamcraft}.

A large body of work has been focused on creation of assets that could be used in games. Much of this focus has been on the automated creation of 3D assets and textures, for example \cite{richardson2023texture, chen2023text2tex, liu20233dall, faruqi2023style2fab, tochilkin2024triposr}.
Text2AC \cite{text2ac} is a hybrid approach using LLMs and diffusion models to generate game-ready 2D characters. Automated animation of rigged assets has been considered for example in \cite{huang2023real} using LLMs to generate structured outputs of joint rotations at key frames. Animation generation has also been explored in \cite{10.1145/3641520.3665309}.


For a general overview of language models in games, including in PCG, see the surveys on the subject~\cite{yang2024gpt, gallotta2024large}. Perceptions of game developers on the usage of generative AI tools in game design have been recently studied in \cite{resistanceisfutile, goldenera}.





\section{Methods}
\subsection{Design goals}

Our system has two primary design objectives.
The first is to push the capabilities of large models in autonomously generating interpretable, designer-editable representations of functional game simulations from a single high-level prompt.
That is to say, we seek a system that---rather than replacing existing game development pipelines wholesale---can plug into it by generating code, meshes, assets, layouts (and perhaps in the future, textures, animations, Finite State Machines for NPC AI): the stuff of game design itself.

Second, given a system capable of such output, we seek novel means of exposing its inner working to users, developing new representations via which to control generated content, complementing the established ones and operating at a higher level of abstraction with respect to them.
To validate our efforts along these lines, we conduct an in-person usability study in which participants with interest or experience in game design and generative AI interact with and give feedback on the system.

In order to effectively generate entire simulations from a single prompt, we argue that three elements are crucial. First, just as most game design involves multiple specialists---or at least the donning of many caps by a generalist designer---multiple specialized AI agents are required to competently perform the plethora of semi-independent tasks involved in full-fledged game design. 
In our case, both are present, because while there are specialists for asset generation and downloading,  most of the others amount to different system prompts for a single LLM, with prompts populated dynamically with the flow of information through system, namely (user or system-generated) instructions, intermediary code and assets, and engine feedback. 

Second, these specialists must be orchestrated dynamically and automatically, as not all game design endeavors involve the execution of the same tasks in the same order.
In the case of our cow abduction example, the planning module ultimately determines what the flow of information (i.e. the order of implementation steps) between implementation submodules will look like.
The system may elect to first use placeholder assets, or to first implement a rudimentary version of cow pathfinding, which, after addition of the spaceship, is extended to include a beam-avoidance mechanic. In this way, the planning module is responsible for effectively untangling the web of dependencies between entities in a complex simulation so that it can be attacked piecemeal.

Finally, this orchestration must be carried out via hierarchical planning, with high level goals broken down recursively into sub-goals and implementation tasks to ensure coherence between high-level objectives and actual implementation steps.


In terms of the user interface, we develop a node-based visual interface for editing steps in the hierarchical planning tree, pruning or expanding sub-plans, and providing feedback on the simulation resulting from generated code and assets.



\subsection{System overview}



    
   

In this section we give a detailed description of the DreamGarden pipeline.
We will describe the planning module along with the implementation submodules which are invoked to carry out this plan.
Unless mentioned otherwise, calls to these modules involve dynamically prompting an LLM, asking for structured output following a given format, and parsing out elements of the response using regular expressions.

The heaviest lifting for which this LLM is responsible ultimately comes via the \textit{code generator}, detailed below. We note that code generation models like \cite{Zhao2024CodeGemmaOC, Lozhkov2024StarCoder2A, qwen} are trained on large volumes of C++ code which grant them capabilities in implementing a wide range of functions in Unreal Engine. However, these models still require precise prompting and a well-structured system architecture to generate fully functional environments within Unreal. In our experiments, we use GPT-4o~\cite{gpt4o} for its speed and versatility.

\subsubsection{Planning module}

We begin with the planning module, whose input is the seed prompt, which is open-ended and may describe a dream, an imagined scenario, or a rough game sketch.
This is fed into the \textit{broad planner}, which is system-prompted to take this open-ended input and reformulate it as an outline for a video game to be implemented in Unreal Engine.
This submodule is also asked to create a broad, high-level plan for implementing the game.

In the case of our cow abduction example, given the initial prompt (see Supp. Mat. for full prompts), the planner responds with the following:

\begin{lstlisting}[label={broad_plan_example},caption={Example broad planner output.}]
Game Conceptualization:
The game is a zero-player simulation where a spaceship flies around the sky, periodically descending to abduct cows and other animals from an enclosure behind a farm. The aesthetic will be whimsical and light-hearted, with cartoonish graphics and humorous animations. The camera will be positioned to provide a clear view of the farm, enclosure, and spaceship's movements. The primary focus will be on the interactions between the spaceship and the animals, creating an entertaining spectacle for the viewer.

Game Design Plan:
1. Design the environment layout
2. Implement spaceship behavior
3. Design animal assets and their interactions
\end{lstlisting}

This high level game description and broad plan is parsed with a regular expression, with text under \verb|Game Conceptualization| converted to the root node in the plan tree, and the numbered bullet points under \verb|Game Design Plan| converted to its child nodes.

Next the \textit{sub-planner} is called recursively to grow this plan into a tree.
This submodule is shown the entire plan tree in its current state, and asked to re-articulate a particular node (a step or sub-step of the plan) in more detail, then further expand it into a list of sub-steps. 

For example, in our cow-abduction scenario, given, e.g. the sub-planner prompt for the second child node in the above broad plan, the model responds:
\begin{lstlisting}[label={lst:subplan_example},caption={Example sub-planner output}]
Node Description: Implement spaceship behavior
Sub-Steps:
1. Generate spaceship asset. LEAF. DIFFUSION MESH GENERATION.
2. Implement spaceship flying behavior, including periodic descent.
3. Implement animal abduction mechanics.
\end{lstlisting}
The numbered bullet points in this response then become their own child nodes.

This process repeats, with the sub-planner processing nodes of the tree in breadth-first order.
The sub-planner is able to flag a generated node as a leaf node (e.g. the first child in the above response) by appending special text.
This indicates that the child node in question is a leaf: the sub-planner will not generate any further children for it while traversing the tree.
The sub-planner's tree generation process terminates when all plan nodes have children or are marked as leaf nodes.

Both planner sub-modules are given a description of the available implementation submodules---each specializing in a particular concrete code or asset generation task---and instructed that the tree they generate must ultimately terminate in leaf nodes that each correspond to an individual task for one of these submodules.

The final hierarchical plan in our running example is as follows:
\begin{lstlisting}[]
1.  Design the environment layout
  1.1.  Generate the farm environment asset. LEAF. DIFFUSION MESH GENERATION.
  1.2.  Generate the animal enclosure asset. LEAF. DIFFUSION MESH GENERATION.
  1.3.  Arrange the environment in the scene. LEAF. ACTOR GENERATION.
2.  Implement spaceship behavior
  2.1.  Generate spaceship asset. LEAF. DIFFUSION MESH GENERATION.
  2.2.  Implement spaceship flying behavior, including periodic descent.
    2.2.1.  Implement spaceship initial flight path and movement logic. LEAF. ACTOR GENERATION.
    2.2.2.  Implement periodic descent behavior for the spaceship. LEAF. ACTOR GENERATION.
    2.2.3.  Implement randomized flight patterns and altitude changes. LEAF. ACTOR GENERATION.
  2.3.  Implement animal abduction mechanics.
    2.3.1.  Implement logic for detecting animals within the spaceship's abduction range. LEAF. ACTOR GENERATION.
    2.3.2.  Implement animation and effects for the abduction process (e.g., beam effect). LEAF. ACTOR GENERATION.
    2.3.3.  Implement behavior for removing the abducted animals from the scene. LEAF. ACTOR GENERATION.
3.  Design animal assets and their interactions
  3.1.  Generate cow asset. LEAF. DIFFUSION MESH GENERATION.
  3.2.  Generate other animal assets (e.g., sheep, pigs, chickens). LEAF. DIFFUSION MESH GENERATION.
  3.3.  Implement animal behavior and animations. LEAF. ACTOR GENERATION. 
\end{lstlisting}

The leaf nodes are then translated into prompts or other formatted inputs for the implementation submodules by the \textit{task generator}, whose prompt is adapted to give general guidance on how best to prompt for the particular implementation submodule at hand.

Child node 2.1 in the sub-planner output in \autoref{lst:subplan_example} for example is passed to the \textit{task generator}, with the model responding:
\begin{lstlisting}[label={task_example},caption={Example task generator output.}]
ASSET NAME: Spaceship
IMAGE PROMPT: Generate an image of a detailed spaceship asset with a sleek, futuristic design. Include smooth metallic surfaces with sharp edges, multiple thrusters at the rear, and an illuminated cockpit at the front. Incorporate advanced technological elements such as antennas and sensors. The spaceship should have a streamlined, aerodynamic shape, with wing-like structures extending from the sides. Ensure the scene is well-lit to highlight textures and details. The image should be without a background, high resolution, and suitable for use in Unreal Engine.
\end{lstlisting}
which amounts to a task for an implementation submodule (the diffusion mesh generator in this case, discussed below).

Once the \textit{task generator} has transformed all leaf nodes into implementation tasks, these tasks are executed in sequence---i.e. visiting leaf sub-plan nodes in order---with the generated code, assets and layouts from prior tasks fed as input to subsequent submodules. 
This process of recursively growing the plan tree then translating its leaf nodes into implementation task prompts is shown schematically in~\autoref{fig:plan_tree}.

\subsubsection{Implementation submodules}


\paragraph{Code generator}

\begin{figure*}
\includegraphics[width=.6\textwidth]{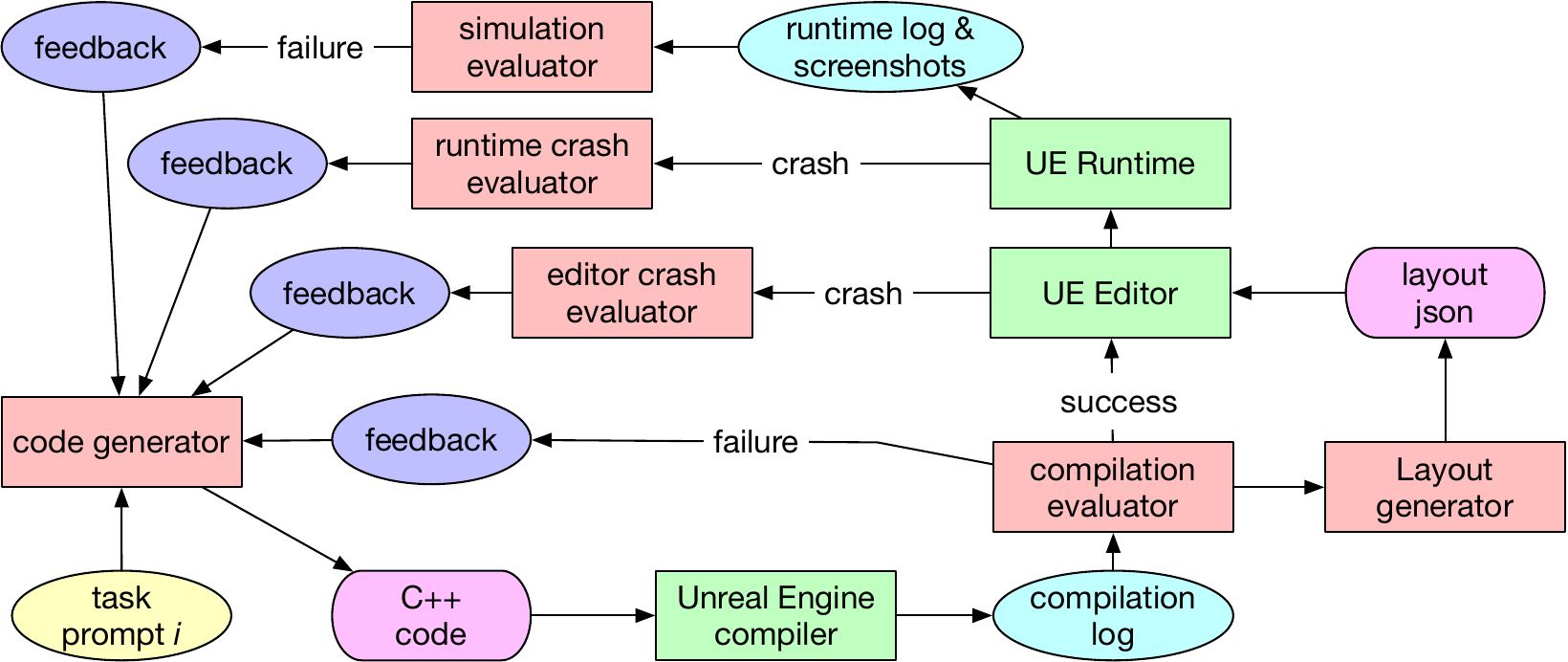}
\caption{The code generation submodule pipeline. C++ code for Actors in Unreal Engine is generated given an implementation task prompt. Feedback is generated given any errors resulting from compilation, python editor scripting, or runtime.}
\label{fig:code-feedback}
\end{figure*}

The coding submodule is the most general and complex submodule in DreamGarden (see \autoref{fig:code-feedback}).
It involves the generation of C++ code for Unreal Engine ``Actors''---essentially, general discrete entities within the simulation capable of arbitrary behavior and interactions---and the generation of an initial layout for these actors in the level, which is then instantiated via a hard-coded python script inside the editor. 
While \textit{DreamGarden} is capable of creating single-player games, we mostly restrict it to 0-player simulations, so that it can provide self-contained feedback to itself without the need for a generalist game-playing agent, such as SIMA \cite{raad2024scaling}.

The task prompt for the coding submodule involves distinct actor and spawner prompts.
First, the actor prompt is fed to a code-generator LLM, whose system prompt contains some directives on writing C++ actors for Unreal Engine, and lists of all materials and meshes available in the Starter Content directory, as well as a list of any meshes from asset downloading/generation tasks and code from any previous code generation tasks that have already been carried out. 
A set of complete C++ files are parsed out from this model's output.

These files are then copied to the UE project directory, and the project is re-compiled.
The compilation log (along with the generated code and task implementation prompt) is fed to a compilation evaluator submodule, which outputs a verdict of success or failure, providing feedback on the source of the compilation errors and guidance on how to resolve them in the latter case.
If compilation is successful, then a spawn layout generator LLM is queried for a layout which places instances of generated actors in the scene by specifying their coordinates, scale, rotation, and initial values for any editable properties of these actor classes in the generated code (note again that some actors might not be visual in nature).
This LLM is shown the generated code, its portion of the implementation task prompt, and some relevant pointers such as the correspondence of UE units to centimeters.
A layout json is parsed out from the output of this model, and the UE Editor is launched.
A hard-coded python initialization script is run when the level is loaded, which places instances of the generated actor classes in the scene.
Should this python script error out---in particular if the layout generator has referenced a class that does not exist in the project's generated C++ code---then the relevant section of the UE log, the initial code, and task prompt are fed to an LLM to produce feedback and pointers on how to address the issue.
Supposing the python script does not raise any exceptions, then actor instances are initialized in the scene and simulation code is launched.
If the editor crashes due to runtime errors, the crash log is again fed to an LLM alongside code, layout and task prompt to produce feedback and guidance.
If the simulation runs successfully, then a hard-coded `FilmCamera' class (protected from being overwritten by generated classes) captures 6 screenshots of the first 6 seconds of in-engine simulation\footnote{In an ideal world we would like to pass more frames from a longer video clip, but that would significantly increase the number of input tokens, in some cases beyond what can fit in GPT-4o attention window.}, and these screenshots along with the runtime log, actor code, layout, and task prompt are fed to a Vision Language Model (VLM) which must then output a success/fail verdict, and provide feedback in the latter case.

Supposing failure at any of the above steps, the resulting feedback, along with the generated (broken) code and layout, and the task prompt, are fed back into the code generator, repeating this process for a fixed number of attempts before the system is forcibly moved on to the next implementation task.

Regardless of success, output code and feedback is injected into the prompt for subsequent coding submodules (see Supp. Mat. for a full example prompt).

\paragraph{Procedural mesh generator.} A variant of the coding submodule, the procedural mesh generation submodule, involves modifying the coding submodule's system prompt to request code for the generation of a procedural mesh for a given structure or terrain.
An example of a procedural cube mesh written inside a C++ UE actor is included in the prompt for in-context learning.
If the code for the procedural mesh compiles successfully, a default layout is generated, wherein a single instance of the actor in question is spawned at the origin for visual examination.
The same feedback loops, with a fixed number of attempts, are applied in this submodule.
In \autoref{fig:terrain_iteration} we show outputs resulting from the chaining together of two procedural mesh generation tasks, with the first generating hilly terrain, and the second adding vegetation, which it adjusts to make more realistic and diverse after visual feedback.

\begin{figure*}
\begin{subfigure}{.45\linewidth}
\centering
\includegraphics[width=1.0\linewidth]{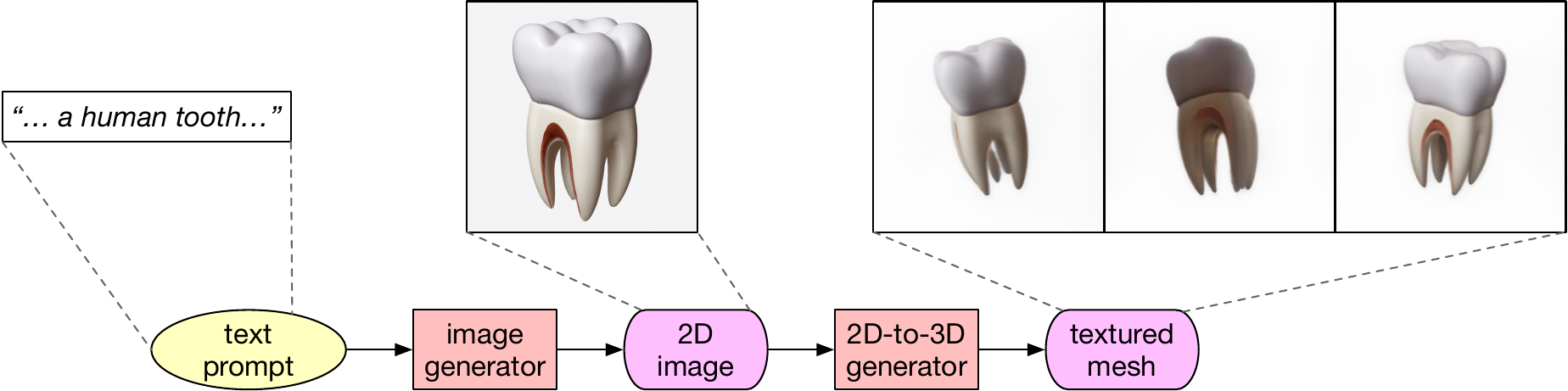}
\caption{The diffusion mesh generator submodule chains together a text-to-2D with a 2D-to-3D model to generate novel textured meshes from text prompts.}
\label{fig:diffusion_mesh}
\end{subfigure}
\hfill
\begin{subfigure}{.45\linewidth}
\centering
\includegraphics[width=1.0\linewidth]{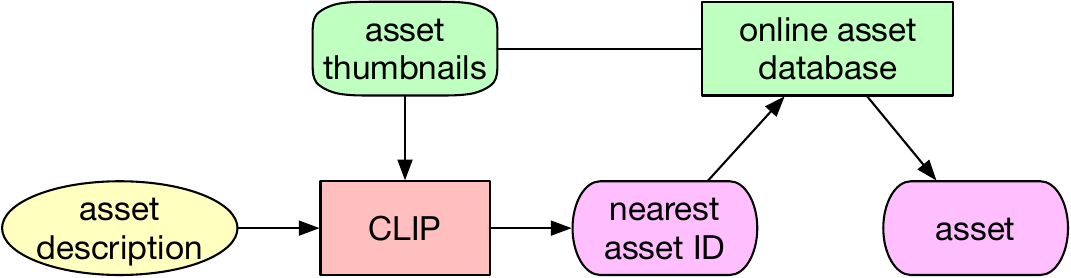}
\caption{The asset downloader submodule compares a concise text description against a large database of asset thumbnails via a contrastive text-image embedding model to find appropriate textured meshes from a database of human-generated content of common objects.}
\label{fig:mesh_download}
\end{subfigure}
\caption{Submodules for generating/downloading meshes.}
\end{figure*}

\paragraph{Diffusion mesh generator.} The diffusion mesh generation submodule~(\autoref{fig:diffusion_mesh}) generates novel 3D assets from scratch, by chaining together a text-to-2D diffusion-based image generator and a 2D-to-3D textured mesh generator.
The task prompt for this submodule is a detailed description of the desired asset.
The initial prompt is appended with some further directives in an attempt to make the image a suitable prompt for a 2D-3D model (i.e. requiring the object to be fully visible and posed against a blank background).
This description is fed into a text-to-2D image generator, which we chose to be Dall-E 3 \cite{dalle3} in our experiments.
The resulting image is then fed into a 2D-to-3D generator, in this case TripoSR~\cite{tochilkin2024triposr}, which we chose due to its speed and SOTA quality at the time of experiments, but this is modular and could be swapped to a completely different model.

The resulting 3D object file is then converted into an UE asset in the engine, and added to a folder within the game's project, to be later exposed to coding submodules for use in the scene via generated actors.

\paragraph{Mesh downloader.} The mesh download submodule~(\autoref{fig:mesh_download}) uses text descriptions to retrieve and import handmade assets from a pre-existing database.
For this submodule, the task generator produces a concise description of the desired asset.
This text is then embedded via CLIP~\cite{radford2021learning}.
The CLIP embeddings of uploaded thumbnails of all assets in the Objaverse database~\cite{deitke2024objaverse} are first precomputed. Then, the distance between these thumbnails and the generated text description are computed, and the the asset whose thumbnail has least CLIP distance to the text description is downloaded from the database.

Both the asset downloader and diffusion mesh generation submodules launch UE in the final step, and, via a hardcoded python script, import the resultant GLB or gLTF asset files into UE, converting them to textured meshes within the engine and saving them to disk for later use.

\subsubsection{Graphical user interface}

\begin{figure*}
\begin{subfigure}{0.48\linewidth}
\centering
\includegraphics[width=1.0\linewidth]{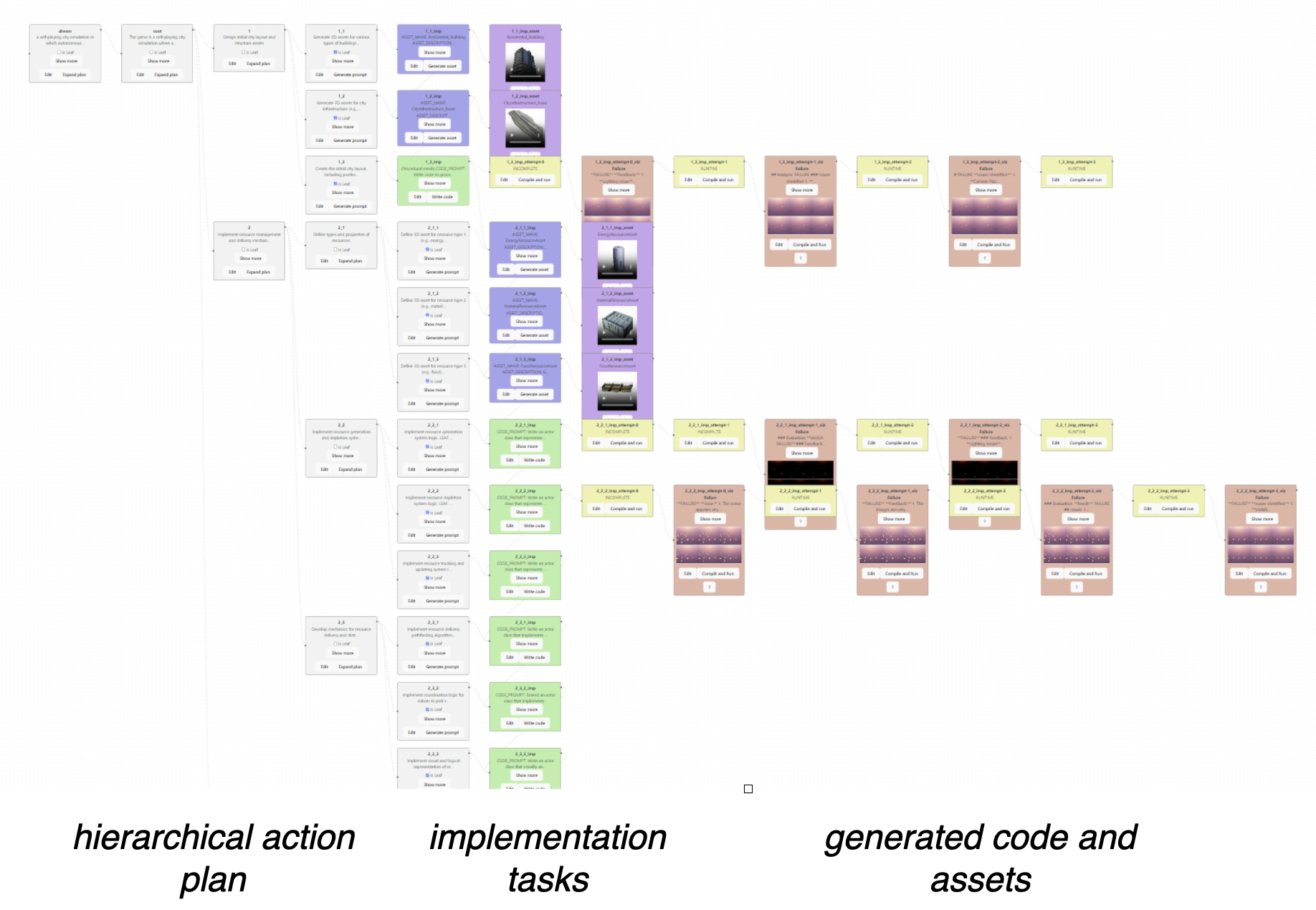}
\caption{Users interact with DreamGarden via a node-based Graphical User Interface, pictured here after the generation of the plan tree (left), implementation tasks (middle), and during iteration---involving code generation, in-editor simulation, automatic visual feedback, and code re-generation---on these implementation tasks by specialized submodules (right).}
\label{fig:gui}
\end{subfigure}
\hfill
\begin{subfigure}{0.48\linewidth}
\includegraphics[width=1.0\linewidth]{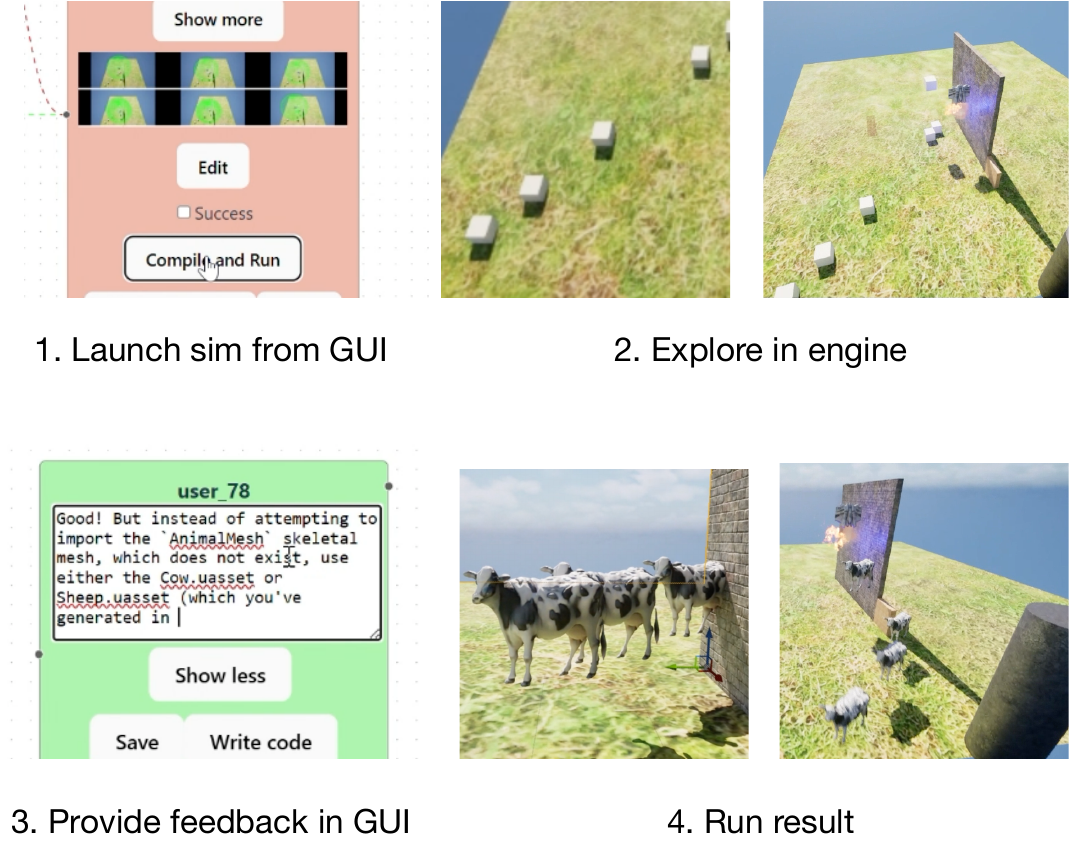} 
\caption{Via the Graphical User Interface, users can compile and launch the generated simulation at various stages of development, exploring the result inside the Unreal Engine editor. They can then provide customized feedback to the code generator.}
\label{fig:user-feedback}
\end{subfigure}
\caption{Users can seed the autonomous growth of a tree of plan and actions (left), and provide detailed feedback by engaging with generated simulations (right)}
\end{figure*}

We develop a Graphical User Interface (GUI) for \textit{DreamGarden}, to expose its growing the user's seed prompt in a more intuitive and visually informative way.
Because the action plan is effectively a tree, the basis of our GUI is a node-based editor, i.e. an interface in which users can manipulate nodes in a tree.
Nodes are either (sub-)steps in the hierarchical plan, implementation tasks, or steps taken by implementation submodules, and edges correspond to the hierarchical relationship between planning steps and tasks, and the the chronological flow of implementation steps.
The GUI is shown in \autoref{fig:gui}

To facilitate responsiveness between the system's backend and the GUI, we structure the backend call as a repeated function on a ``frontier'' of nodes to be expanded/implemented.
The user's edits can effectively result in pruning this tree, modifying nodes, and adding/removing them from this frontier.

The workflow is as follows.
The user begins with an empty canvas, and can add a seed node, where they enter the open-ended high level prompt.
They then have the option to ``step'' or ``play'' the system, either stepping through a single call to the backend node-expansion function, or initiating the repeated expansion of nodes.
(The inclusion of the step function is crucial in cases where users intend to take a more active role: if instead the backend is set to play, then user edits on nodes that have already been expanded may result in the computationally costly expansion of nodes which are ultimately pruned as a result of user intervention.)

The seed node is expanded first into a broad plan, with children nodes containing plan steps fanning out to the right.
This kind of expansion is applied recursively to the tree, until all extant nodes have children or are leaves.
The placement of nodes can be adjusted, and they can be expanded to show the full text generated by the system which describes the plan step they represent.

Once the plan tree is fully expanded, the leaf nodes of the tree are added to the frontier and queued for transformation into task prompts.
That is, the task generator will iterate through these leaves in order, transforming them into a specific input for the submodule to which each task is assigned, represented in a new node to the right of its corresponding leaf plan node.
While all plan nodes are represented in neutral light grey, these task nodes are color-coded according to the submodule to which they correspond,
see \autoref{fig:gui}.

\paragraph{User edits}

Each node in the action tree has an ``is leaf'' checkbox.
When toggled by the user, this determines whether or not the given node is considered as a leaf in the tree (i.e. the description of a concrete implementation task to be handed off to a particular submodule).
By turning a non-leaf into a leaf node (i.e. when the system is unnecessarily granular), the user effectively prunes the tree.
During the next call to the node-expansion function, any plan nodes that are children of this node are removed, as are any task nodes attached to these children.
Similarly, if these tasks have already resulted in implementation step nodes, then these implementation nodes are also removed (this data is backed up to a separate folder, should the user want to reverse their changes).
Finally, because any code and assets resulting from these implementation steps would have been handed down to implementation tasks following it in the ordered sequence of leaf nodes, then any implementation steps belonging to these tasks are also backed up and deleted.

By turning a leaf node into a non-leaf node, the user effectively adds this node to the frontier, queuing it for expansion into a sub-plan.
Again, if a task node was generated for this leaf, it is deleted, along with any implementation nodes belonging to this task, or any implementation nodes belonging to tasks following it in the ordered list of tasks.

During code (and procedural mesh code) generation, coding attempts and code evaluations are displayed in nodes that are chained in a sequence, moving from left to right out from the parent task node (itself stepping from a leaf node of the plan tree).
When the system produces code that compiles and runs successfully in Unreal, screenshots are automatically captured of the first moments of simulation.
These screenshots, along with feedback produced by a vision language model based on them (in addition to the runtime logs and generated code and actor layout), are displayed in a special ``visual evaluation node'', which is chained to the ``code attempt'' node from which it stemmed.
These nodes are augmented with a ``compile \& run'' button, which allows the user to compile the code as it was at this state, and launch the editor with the corresponding actor layout.
The user is then able to inspect the resultant simulation in Unreal Engine.
They may, for example, make edits to generated code (testing them via recompilation from within the editor), which when saved would then be fed into subsequent iterations of code generation.
They may also return to the GUI, and edit the feedback generated by automatic evaluation.
This will then result in the pruning of all implementation steps having this node as parent (or, spatially speaking in the GUI, to the right of or below it), and the rerunning of the subsequent code generation iteration following it.
In \autoref{fig:user-feedback}, a user launches an intermediate simulation from our running example.
They find the code in a state wherein the cows are wrongly not loading the generated assets.
Entering this feedback into the relevant GUI node and resuming the system results in a subsequent node where the cubes are replaced with realistic cow meshes.

\subsection{Study design}

To explore the interaction affordances of our system, validate it against user needs and behaviors, and explore directions for future development, we conducted a usability study with $N=10$ participants. These participants were  recruited via emails and posts in messaging groups from within the researchers' organization.
Users first filled out an intake survey, in which they specified their levels of experience with game design, generative AI, and Unreal Engine.
The users were then given a brief overview of the system's intended purpose, and the means they had available to interact with it.
Users were first invited to imagine a high-level seed prompt, with little restriction on form or content, which they then entered into the initial ``dream'' node in the GUI.
The study was conducted in person, with the DreamGarden system running on a laptop with a top-tier GPU.
Users were encouraged to observe the system's operation by interacting with the GUI, as it expanded their prompt into an increasingly detailed plan of action, assessing its viability by talking through it out loud.
They had the option of pruning or expanding the tree further by toggling leaf node status via the GUI.
They were then invited to observe the operation of the implementation submodules; examining assets in the GUI or a separate file explorer window, and assessing generated code files in a separate IDE window if desired, and to the best of their ability.
Users were also reminded at this point that, as per the consent form they had signed prior, they were free to check their phone, grab a drink, or otherwise disengage from the system's operation as it grew artifacts autonomously.
When the system generated code that compiled and ran successfully in-engine, they were invited to observe visual feedback nodes, compile and run the code at points that seemed interesting, explore the resulting environment via the UE Editor and provide customized feedback in the corresponding GUI nodes based on their first-hand observations. Finally, after using  DreamGarden, the participants were asked to answer a series of questions in the exit survey on their experiences with it, including rating their frustration and enjoyment of using the system on on a 5-point Likert scale (1 = Strongly disagree, 5 = Strongly agree). See the Supplementary Material for the detailed questions.
Studies lasted a maximum of 45 minutes each.
The participants were invited to submit seed prompts ahead of arrival, allowing them to step ahead at their own pace through the generation of a partially complete garden without having to wait for the system to catch up.

The user study design, including calls for participation, the structure of the in-person study, the user workflow with respect to the system, and the intake and exit survey questions, was approved by an Internal and Ethics Review Board prior to conducting the study.

\subsubsection{Data collection and analysis}

In addition to the intake and exit surveys, the researcher present during the study took notes of the participants experience with and commentary on the system.
Additionally, each participant consented to audio and screen recordings of their study session. Audio recordings captured participants' comments during their interaction with the system and discussion surrounding the design process. These recordings were transcribed using Microsoft Teams transcription software. Screen recordings captured interactions with the system (both the GUI as well as the file explorer and code editors used by some users to more closely track the system's output).
Finally, DreamGarden automatically logged the work of the system in concert with the user---including any user edits, and backups in the case of any resultant tree-pruning---that is, the sequence of all prompts, responses, generated code and artifacts, and in-engine screenshots produced over the course of the study.

Following guidelines for qualitative HCI practice~\cite{mcdonald2019reliability}, authors then collectively reviewed notes, transcripts, and system output, identifying recurring topics and behavioral themes to uncover emergent themes pertinent to our research questions~\cite{fielding2012triangulation}.

\section{Results}

\subsection{Qualitative system observations}

We experimented with various hyperparameters, most notably the branching factor $b\in\{1,\ldots,5\}$ and maximum depth $d\in\{1,\ldots,5\}$ of the tree generated by the planning module, and the maximum number of attempts $i\in\{1,\ldots,4\}$ allowed to the coding generator before forcibly moving on to the next task.

\begin{figure*}
\begin{subfigure}{0.47\linewidth}
\includegraphics[width=\linewidth]{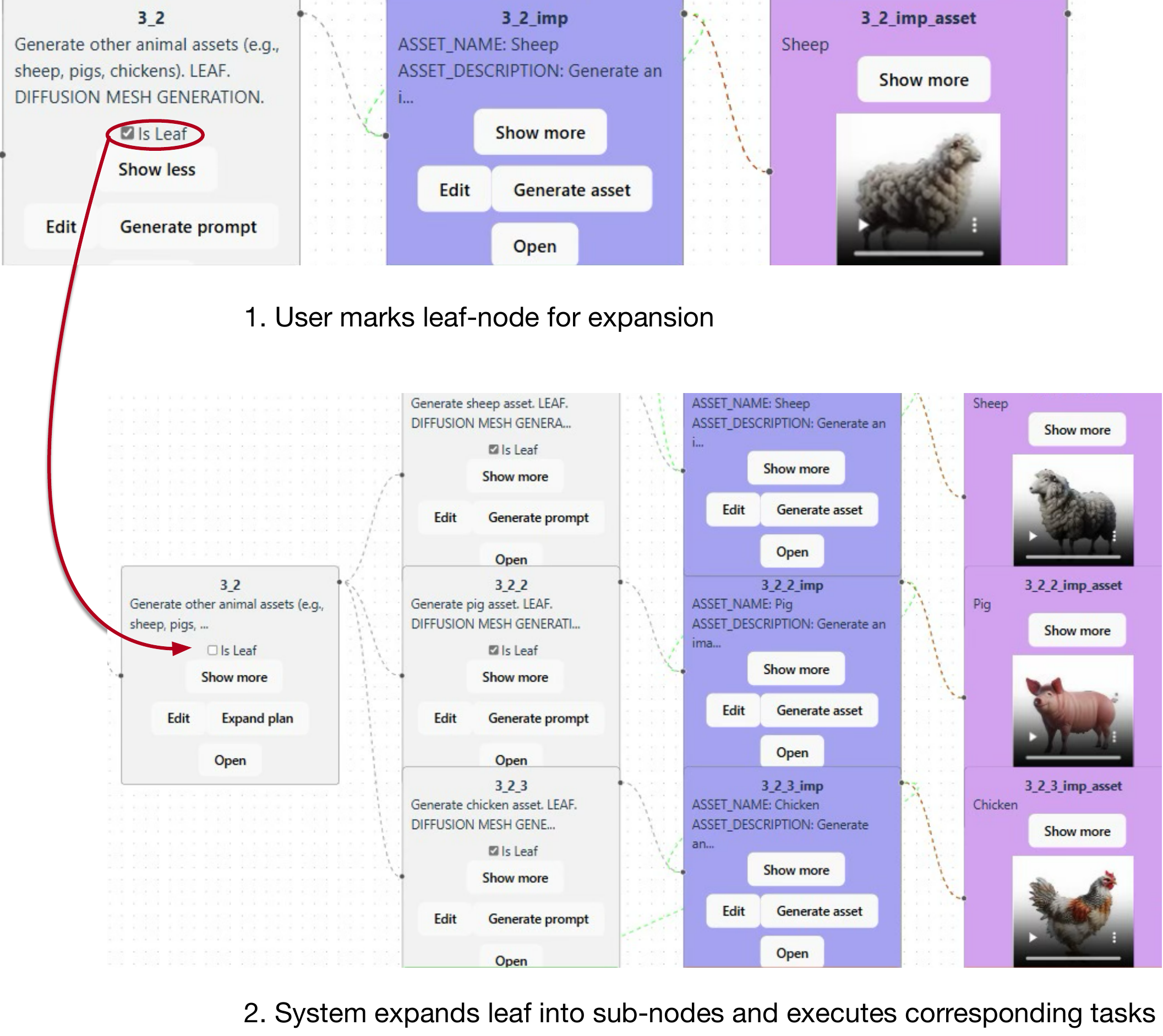}
\caption{By unchecking the `Is Leaf` option, a user can mark a given leaf node for further expansion by the sub-planner. Here the user marks for expansion a that erroneously called for the generation of multiple animal types via a single diffusion mesh generation task. Once the node is expanded, all listed animal types are generated via separate calls.}
\label{fig:enter-label}
\end{subfigure}
\hfill
\begin{subfigure}{0.47\linewidth}
\includegraphics[width=\linewidth]{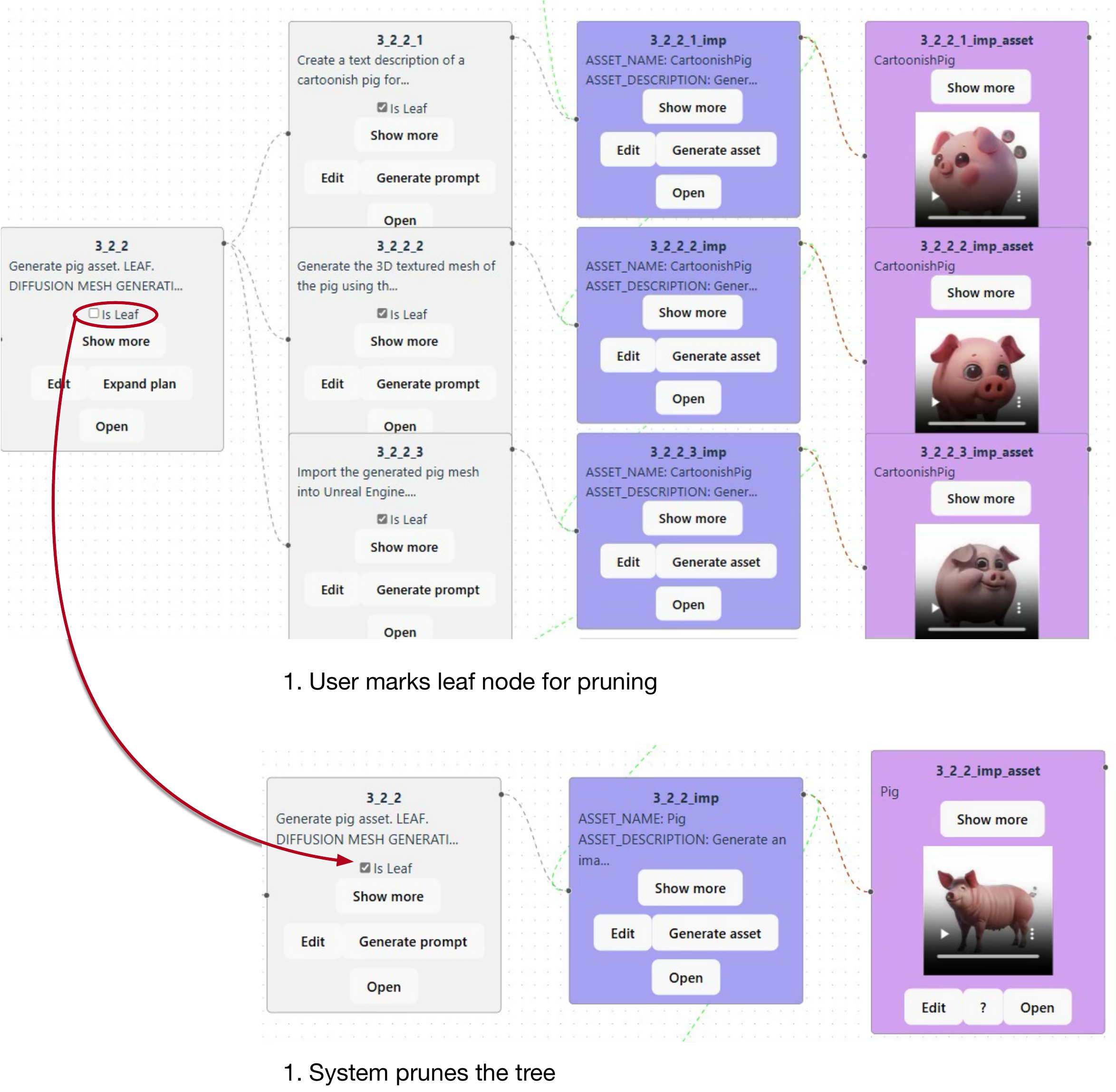}
\caption{By checking the `Is Leaf` option, a user can mark a given leaf node for pruning. Here the user marks for pruning a node that calls for the generation of a single animal asset which has been superfluously expanded into multiple child nodes, resulting in redundant assets. After pruning, a single, standalone asset remains.}
\label{fig:enter-label}
\end{subfigure}
\caption{Users can prune or expand sub-trees by interacting with plan nodes in the GUI.}
\end{figure*}

We note that the system will not always expand the tree to the fullest extent.
Still, in simpler scenarios, like ``a sheep grazing on a grassy hillside'', it will often superfluously subdivide certain steps, e.g. breaking down ``generate sheep asset'' into separate tasks for the generation of individual body parts, or for the generation of mesh and texture.
This latter case speaks to a problem we notice occasionally with the system, in which it will invent tasks beyond the capability of the submodules to which it currently has access, or at least which are not explicitly mentioned to it.
This was the case with lighting, atmosphere effects (such as fog) and skyboxes, for example, where the system would sometimes implement these features despite their conflicting with the level's default lighting.
To remedy this, we  removed all environment assets from the default level, prompted the coding module to account for these explicitly, and gave an example of a sufficient `EnvironmentManager` Actor class in its prompt.
With animation, on the other hand, we adapted our prompts to insist that the system would not be capable of animating skeletal meshes.

When we attempt to restrict the tree to particularly small depths and branching factors, it sometimes ignores these requests, for example adding too many children to the root node.

We note anecdotally that for more complex tasks, larger depths and branching factors allow for better subdivision of implementation tasks, making it easier for the system to progress toward high-level objectives without getting stuck on implementation tasks involving extending the code in very complex ways in one pass.
Instead, it is easier for the system to add small but observable functional changes to the code step by step.

Naturally, allowing the code generation submodule more generation attempts increases the odds of it finding code that compiles and runs, and results in acceptable visual results.
This improvement seems largely attributable to the specific feedback given by evaluation agents, which are often able to make sense of informative compilation and runtime errors and warning messages, suggesting relevant code fixes for future generation attempts.

For the usability study we set all the above hyperparameters to 3 to strike a balance between complexity and generation time.

\subsection{Usability study results}
\begin{figure*}
\centering
\includegraphics[width=1.0\linewidth]{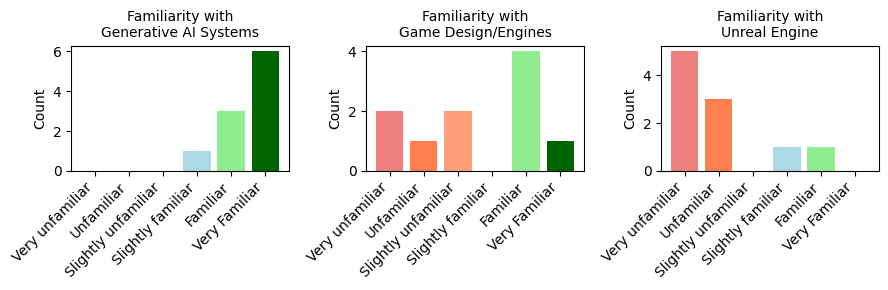}
\caption{All users were at least slightly familiar with generative AI, while they ranged in experience with game design, and were largely unfamiliar with Unreal Engine.}
\Description{Figure 10. User Familiarity with Relevant Domains
This figure presents the distribution of user familiarity levels across three domains:
Generative AI Systems: Most users reported being either "familiar" or "very familiar" with generative AI systems, with a smaller portion slightly familiar or unfamiliar.
Game Design/Engines: User familiarity with game design and engines shows a spread across all levels, with "familiar" being the most common response, followed by fewer users being "very familiar."
Unreal Engine: Familiarity with Unreal Engine skews towards lower levels, with the majority of users reporting being "very unfamiliar" or "unfamiliar."
These insights highlight varying expertise levels among users, which can inform the design and accessibility of the DreamGarden system.}
\label{fig:intake_survey}
\end{figure*}

\begin{figure*}
\centering
\begin{subfigure}[t]{0.49\textwidth}
    \includegraphics[trim={200 200 200 200},clip,width=\textwidth]{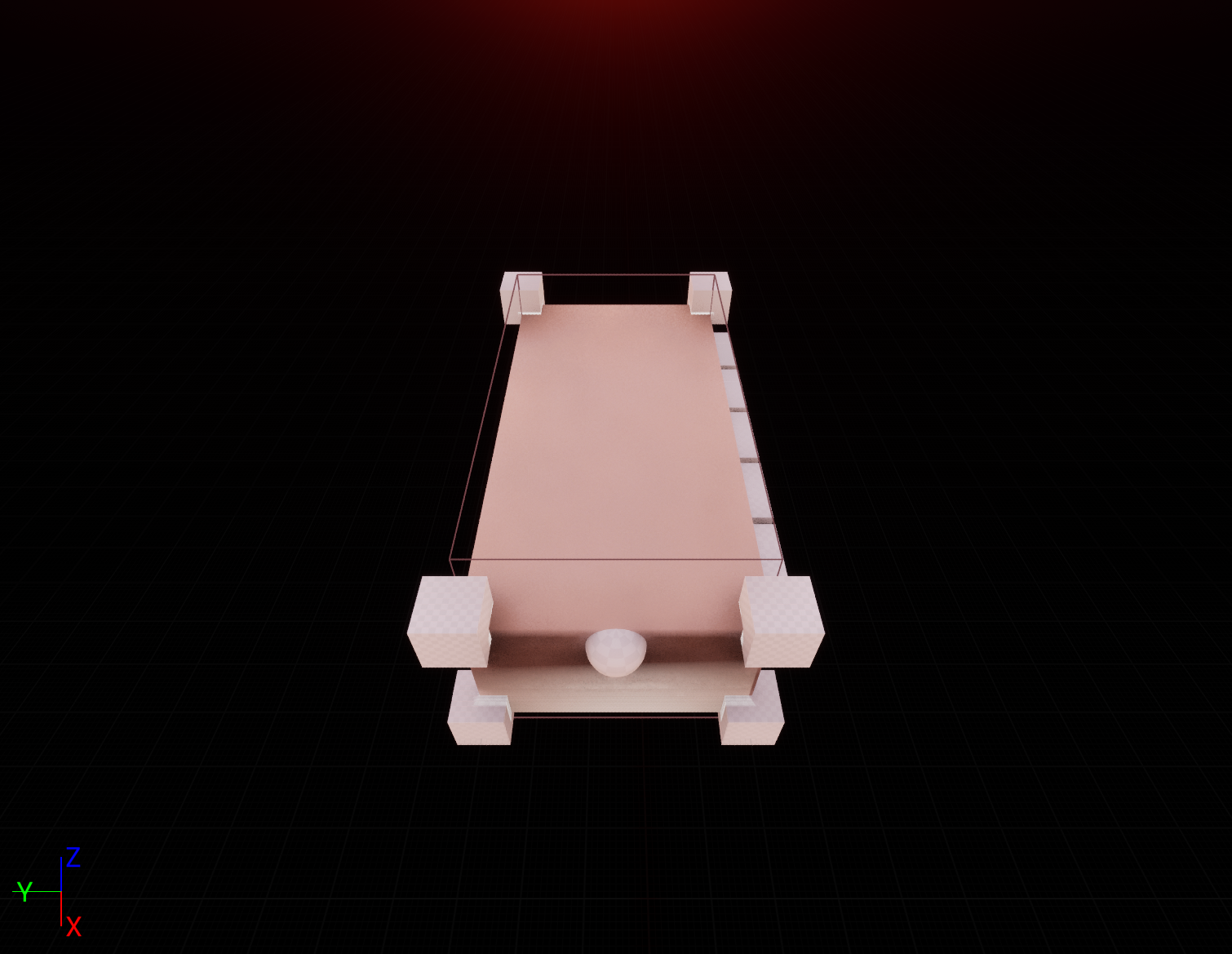}
    \caption{\textit{A classroom -- teeth will be falling out}}
    \label{fig:first}
\end{subfigure}
\hfill
\begin{subfigure}[t]{0.49\textwidth}
    \includegraphics[trim={200 200 200 200},clip,width=\textwidth]{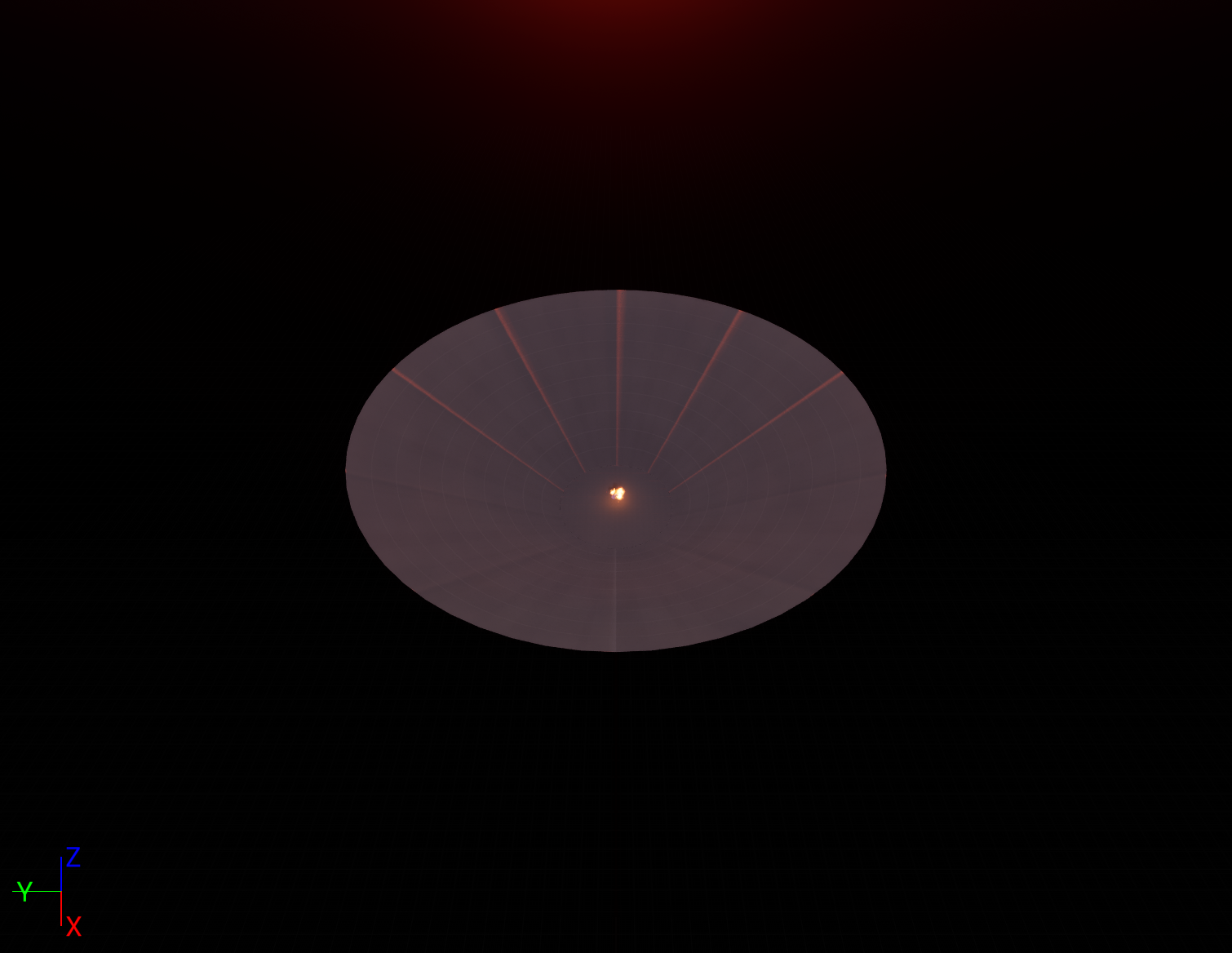}
    \caption{\textit{...The battle arena will be floating in space...laser-shooting unicorns are in the scene...}}
    \label{fig:second}
\end{subfigure}
\hfill
\begin{subfigure}[t]{0.49\textwidth}
    \includegraphics[trim={200 200 200 240},clip,width=\textwidth]{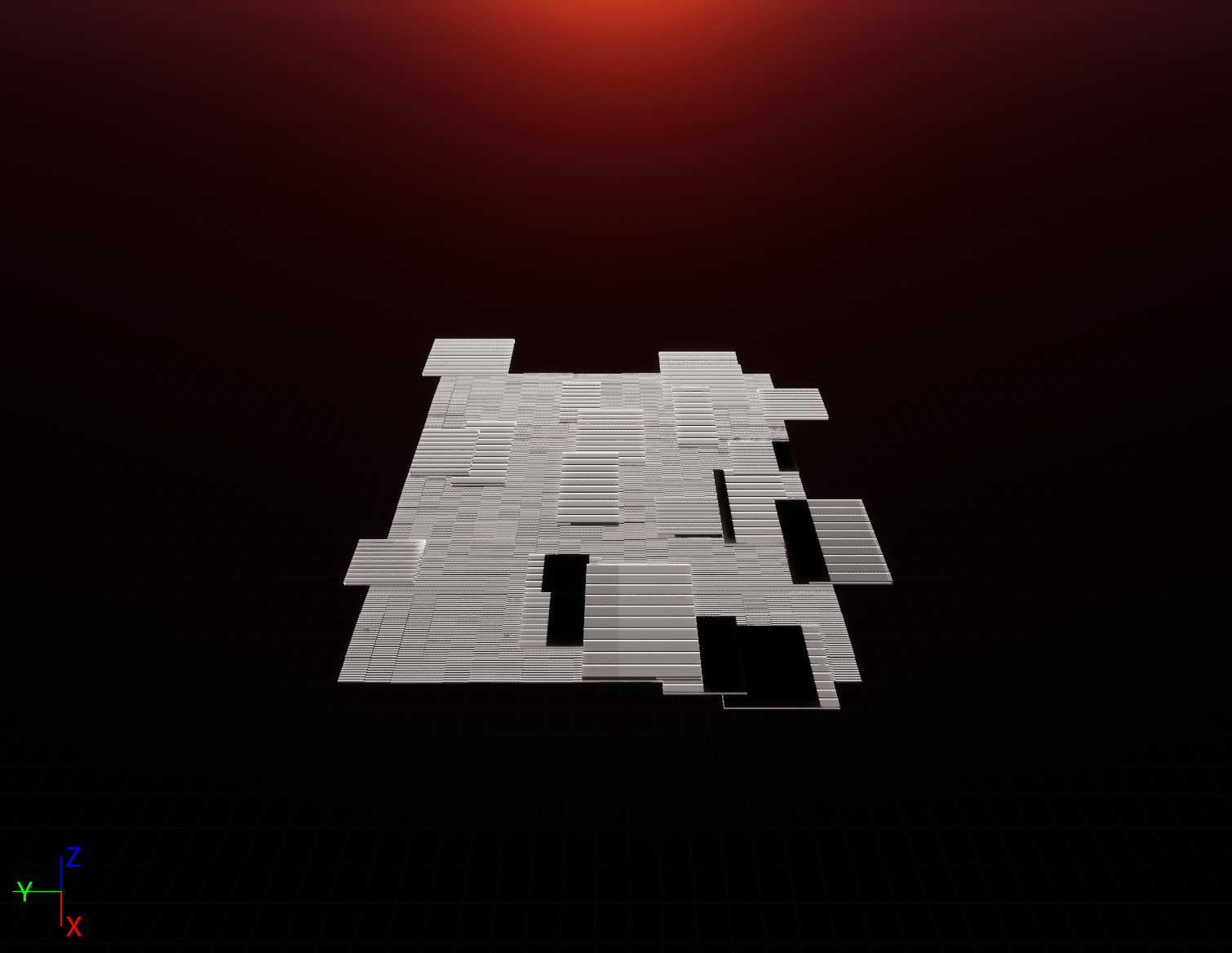}
    \caption{\textit{A platform-like game inspired in the original Mario Bros game but in the style of Minecraft using Steve and only Minecraft characters.}}
    \label{fig:third}
\end{subfigure}
\hfill
\begin{subfigure}[t]{0.49\textwidth}
    \includegraphics[trim={200 200 200 200},clip,width=\textwidth]{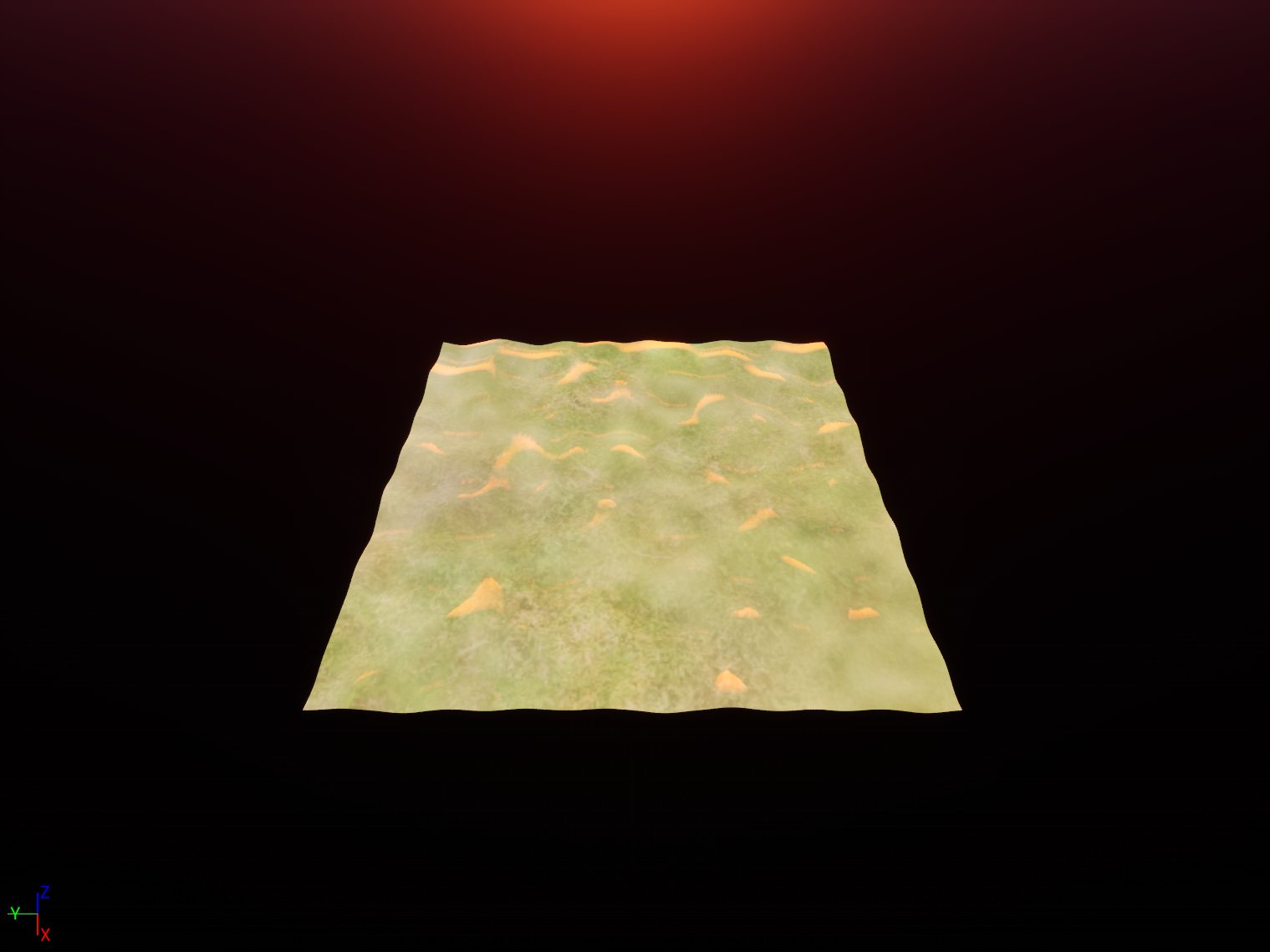}
    \caption{\textit{dragons fighting over a big hill}}
    \label{fig:third}
\end{subfigure}
\caption{Intermediary outputs generated from seed prompts during the user study.}
\label{fig:user_outputs}
\Description{Figure 11. Intermediary Outputs Generated from Seed Prompts
This figure showcases intermediary 3D outputs created by DreamGarden based on user-provided seed prompts during a user study:
(a) A classroom with falling teeth: A rectangular classroom environment, with placeholders representing teeth that might fall during the simulation.
(b) A floating battle arena in space: A circular arena situated in space, designed for a fantastical scene with laser-shooting unicorns.
(c) A Minecraft-inspired platformer: A platform-like layout reminiscent of the original Mario Bros game but styled with Minecraft assets and characters.
(d) Dragons fighting over a hill: A textured hill environment prepared for a scenario involving dragon battles.
These examples demonstrate the system's ability to generate diverse environments aligned with creative and unconventional user prompts.}
\end{figure*}

We will now discuss findings from the user study and group them into a few broad categories. Our survey indicated that all the users were at least partially familiar with generative AI, though their familiarity with game design tools was more varied, while Unreal engine itself was largely unfamiliar to most of the participants, see \autoref{fig:intake_survey}. We present sample intermediary results from the user study generations in  \autoref{fig:user_outputs}.

\paragraph{Utility of planning}
A common theme expressed among users was the utility or potential utility of DreamGarden in the planning and prototyping phases of design, or for creating first drafts of more complex projects.
Indeed, several users expressed that the planning module may have some standalone utility, irrespective of the system's ability to execute these plans itself.
Conversely, users expressed reservations regarding the potential use of the tool in cases demanding more complex game mechanics and environments.
Though not prompted to directly consider the merits of DreamGarden as an educational tool, a natural split on this matter emerged among some users with P10 stating that ``seeing a game design being broken down into steps may be useful to new developers trying to learn how to build their own game'', and P3, on the other hand, stating that they ``would be hesitant about using it for the purpose of really learning, since it takes some of the opportunity to problem solve away from the user''.
Both P3 and P10 were \textit{familiar} with game design and generative AI, and \textit{very unfamiliar} with Unreal Engine.

\paragraph{When not to use such a tool}
When asked to consider what circumstances were \textit{not} befitting of the application of a tool like DreamGarden, several users imagined scenarios in which designers desired higher degrees of control, or more fine-grained control over the final product, with P4 stating ``I would be reluctant to use a tool like LucidDev\footnote{Note that the working title of the system at the time of the creation of the user study survey was ``LucidDev'', and was later changed to \textit{DreamGarden}.} for any kind of deliberately artistic endeavor, as I would prefer to have direct control over the outputs''.
P9 raised the issue of the potential limitations of language, stating that the system would be unsuitable in situations ``where I don't have words to describe what I want''; as well as scenarios in which measures of success were not necessarily visually obvious, or easily described in advance, but rather required ``some intuition to decide if it is good enough''.

\paragraph{Enjoyability}
When asked to discuss elements of the system and experience which they enjoyed or did not enjoy, several users highlighted the pleasure of watching the system grow their seed prompt into a tree-structured plan action, with P8 stating that ``the tree-like visualization of progress and what the system is doing at every step is pretty remarkable'' and P2 stating that they ``found it exciting to track how the system interpreted my simple prompt and how I could see it's plan''.
The fact that this relatively fast pace is not maintained in the GUI (with, e.g., intermediary steps of the coding submodule being hidden from the front-end) seems to have caused some displeasure, with P2 stating that ``the slow process of generating the scene wasn't very enjoyable''.
In this same vein, some users expressed interest in having more intermediate states visualized in the GUI, with P8  suggesting that ``the code output can be shown on the browser window too so you have an even more detailed description of what's going on''.
Some of the intermediary output of the system generated over the course of the user study is shown in \autoref{fig:user_outputs}.

\begin{figure}
\centering
\includegraphics[width=1.0\linewidth]{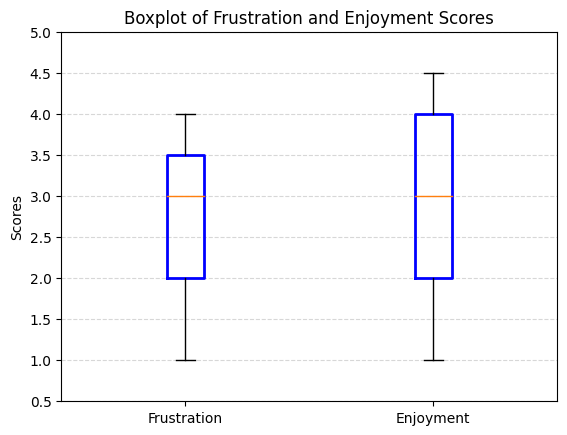}
\caption{Users were relatively balanced in their emotional experiences of the system.}
\Description{Figure 12. Boxplot of Frustration and Enjoyment Scores
This figure compares user-reported frustration and enjoyment scores during their interaction with the DreamGarden system. Both metrics show a relatively balanced distribution, with the median scores (orange lines) positioned near the center of the range. The interquartile range (blue boxes) indicates moderate variability, while the whiskers highlight the full range of responses. Users experienced a mix of both frustration and enjoyment, reflecting the challenges and rewards of using the system.}
\label{fig:user_emotions}
\end{figure}

Users expressed a range of emotions with respect to their experiences with the system~\autoref{fig:user_emotions}.
In their explanations of these ratings, it is clear that expectations and investment play a large role.
While P4 experienced high frustration (4) because the system ``doesn't really work'', they also experienced high enjoyment (4) because ``it was fun to try out and see what it was doing''.
P3, meanwhile experienced low enjoyment (2) and frustration (2), explaining that they were ``not super invested in the outcome''. 

\paragraph{Transparency and Explainability}
In one notable example, P8 requested a ``Super Mario Bros.-like'' platformer ``but with the art style of MineCraft'' in their seed prompt.
However, the system lost the intended MineCraft influence as it sub-divided the plan, generating generic NPC assets instead of stylized ones.
Thanks to the explicit nature of all steps in hierarchical plan generation via the GUI, the user was able to pinpoint the expansion step at which this request was ignored.
This would have made it straightforward to remedy via human intervention.
Cumulative errors in code, on the other hand, were not obvious to users, who generally wound up disoriented and lagging behind system progress when digging through generated code and evaluation---here, output is verbose, dense, and less accessible through the GUI.
Along these lines, several users suggested providing summaries of code changes during generation, or errors during evaluation, with reference to relevant code snippets, making the system's functioning ``easier to digest'' to users, such that they might ``jump in and give it some correction so it doesn't have to flounder so much on its own'' (P7).

\paragraph{Interactivity}
While some users seemed to envision a more generally capable and fully autonomous system, capable of making functional games from scratch without intervention, others imagined ways in which the system might depend on human designers, and even explicitly ask them for help given particularly difficult subtasks or uncertain results.
P8, for example, thinks it ``less important that I give it a simple prompt and come back the next day hoping it did the right thing, and more important that I can craft and sculpt with it directly'', for example ``I could be like "make a big cube in the center" and *poof* a big cube appears''.
P2, meanwhile, thinks ``more iteration on each step would make the end result better'', and that ``this could be human in the loop or not''.

P4, who was \textit{slightly familiar} with Unreal, spent time investigating generated source code, using file editors to examine the sequence of prompts and responses produced by the coding submodule.

Though users were all reminded that in their consent forms, they were welcome to take breaks (check phones, grab drinks) during the study, due to the semi-autonomous and long-timescale nature of the system, only P3 actually left the room to get a drink from a nearby kitchen area.
Several users, on the other hand, initiated some form of small talk with the researcher present while the system iterated on code.

\paragraph{Opinions on the GUI}
P3 pointed out the fact that so-called ``leaf nodes'' in the plan tree are seemingly not actual leaves in the overall graph as displayed in the GUI, given that they themselves appear to be the parents of task nodes.
Similarly, P5 expressed repeated confusion about the togglable ``is leaf'' checkbox available in plan nodes.



Most users focused on the GUI, with several asking for an explanation about the coloring of nodes. Some users (P2, P8) highlighted appreciation for being able to preview diffusion-generated assets in the GUI.
 P3 suggested adding highlighting to the in-progress node, and P3 and P7 both pointed out that coloring nodes red gave a false impression that this node had failed (in the current version of the system, nodes containing screenshots from successfully compiled code that has run in the engine, along with the feedback given based on these screenshots, are colored red, regardless of the overall verdict handed down in this feedback). P3 suggested reserving ``special'' colors like green, yellow and red, for respective states of success or failure.

When the system writes functional code, the editor is ultimately launched to capture screenshots of the simulation at this stage---but for six seconds.
Often, a user would exclaim when, sometimes after a long lull of the system silently iterating on faulty code, an editor window with something in the distance appeared out of the blue.
But of course this window was then automatically killed almost instantly.
This seemed to appear as a somewhat frustrating phenomenon for users, who were sometimes able to begin moving toward the centered object in the simulation in the very last available second.

In terms of hypothetical improvements to be made to the system, some shared desiderata include being able to critique and/or refine assets in the GUI, the ability to edit or delete arbitrary nodes in the garden, and swap out submodule assignment.




\section{Discussion \& Design Implications}

In this section we discuss the findings from the usability study and the implications for design of future versions of assistive systems for game development.

\paragraph{Improving overall performance}
Most obviously, the fact that the system struggled with more complex scenarios, as many users pointed out (or suspected, given the limited time allotted to the interactive portion of the study itself---around 25 minutes---relative to the time actually required by the system to finish iterating on a seed prompt given the default hyperparameters---around 1 hour) points to a general need to make the system more performant overall.
We concede that large models may not be able to produce genuinely novel artifacts relative to the corpus of human-generated content upon which they have been trained.
Nonetheless, by integrating them in a hierarchical, iterative, and specialized pipeline such as \textit{DreamGarden}, we argue that they can at least provide a powerful means of interpolating among this vast space of prior human content, a source of what Kate Compton has termed ``liquid art''~\cite{karth2023conceptual}.
Perhaps the simplest way of making the system a better interpolator of existing artifacts would be to use more thorough few-shot prompting.
Though there is a vast amount of documentation, tutorials and example code snippets pertaining to Unreal Engine online, we have largely refrained from incorporating it into our prompts in the current iteration of the system.
One concern might be where to begin with such a sheer quantity of potential prompting material.
But thanks to the sub-divided and iterative nature of design tasks endeavored by DreamGarden, some obvious opportunities for inclusion emerge.
In particular, during the task generator's creation of submodule prompts, it could additionally select what libraries' documentation should be included in the prompt for the coding submodule with each task, given the description of this task in the leaf plan node, and the names (and perhaps brief descriptions) of high-level C++ modules in UE. Then, the API pages for these libraries could be inserted into the code generator's prompt.
Similarly, when the coding module encounters compiler or runtime errors, the corresponding evaluator LLM might choose to include relevant API documentation in the prompt for the code generator at the following attempt, to rectify any incorrectly called functions or classes.

\paragraph{More digestible presentation}
Regarding the possibility of providing more digestible code snippets, we note that, though in some cases this might involve a tradeoff between interpretability and computational expense---wherein models would have to prompted, after generating or evaluating code, to provide summaries of their work---our generators and evaluators are largely already being chain-of-thought prompted~\cite{wei2022chain} to some degree.
Code evaluators, in particular, usually summarize issues with the code in numbered lists of brief bullet points, which are elsewhere in its response expanded in more detail.
This kind of structured response could be explicitly requested and parsed out to be emphasized in relevant GUI nodes, with the full text expandable via user interaction.
Code generators, meanwhile, sometimes provide brief summaries of their work in natural language before or after the prompt.
We expect that asking for more structured outputs would come at little extra cost to the quality or efficiency of their work and might even make it more robust.

\paragraph{Broader applicability}
The general enjoyment of users had in watching their plans expand suggests that this approach developing high-level into low-level prompts may have broader applicability.
In particular, it may be that for many applications, breaking a high-level goal down into a hierarchical plan might serve to improve the robustness of generated content (at the price of additional compute).
We emphasize that the tree-based planning approach is agnostic to the exact nature of the submodules to be called upon by leaf nodes.
One could thus apply the system to, e.g., robot control problems, the creation of abstract multimedia art, or the generation of a complex narrative.

One might ask how exactly a hierarchical planning tree involving iterative breadth-first expansion via repeated model queries differs from a hierarchical planning tree generated in one pass.
The key difference is that we can control exactly what is provided to the context window during expansion of each node based on said node's placement in the tree, whereas, during one-pass tree generation, any approach will, for example, ultimately expose the last node to all prior nodes, because of the linear nature of the text being generated\footnote{We liken this, by analogy, to the ``graph convolutions'' in neural network architecture~\cite{wu2020comprehensive}, which structure the signals passed to neurons based on a graph structure.}.

Here, more generally, we see hints of a possible alternative to the standard, linear ``chat'' interface, where we instead represent the open-ended generation of text as a graph, where this graph serves to visualize what generated text ought to influence each new generation, with users recombining fragments of text in a way that is more free-form and intuitive than that afforded by strictly linear walls of text.
Users' pleasure in interacting with DreamGarden's node-based GUI would seem to bode well for such a vision.

Especially for open-ended, creative tasks, it would seem that a graph-based approach, resembling the classic ``mind-map'', involving an ever-enlarging frontier of ideas moving out across an infinite plane, serves as a better symbolic representation of AI-augmented, human-driven discovery.
Though AI can make ``liquid'' via sophisticated interpolation the space of prior human expression, by incorporating humans in the open-ended loop one can begin to probe the frontiers of this space, expanding and reshaping it, stretching it to fit new, novel artifacts.


\paragraph{Question of control}
Regarding the concern about degrees of control in situations where the author views themselves as ``auteur'', we note that in theory, the system could play a much more passive role than it does currently.
For example, the system could be paused by default.
A designer could then produce any number of C++ actor files, assets, and create a scene layout, only relinquishing control to the system when, for example, they wanted to receive feedback on how to address a compilation or runtime issue.
Similarly, the user could provide a partially complete project, and experiment with ``modding it'', by giving over control for subsequent tasks intended to extend the current project.
Or, conversely---and as was a common suggestion among users---the system might be used to generate a rough first draft or prototype which is then taken over entirely by a human designer.

Though this promise---of a system with, effectively, a knob that adjusts the degree to which it is autonomous vs. reactive---is seemingly grand, and risks stretching itself thin from a design standpoint by aiming to please too many users too much of the time, it should be relatively easy to deliver if we can ensure that the system is able to observe all edits made by the user, and, conversely, the user is able to observe all edits made by the system.
It need not necessarily be the case the the assistant and the user have exactly the same means of expression, but rather that any action on the part of either can be entered into the conversation.
This is sufficient to allow the system to transition from complete autonomy to complete reactivity, because it ensures that no matter whether user or system take over completely for any amount of time, the other is always able to observe and respond to the changes that have occurred over this period.
To this end, the main focus moving forward with the development of the system should be twofold: both to increase the legibility of the system to the user, for example by following some of the practical suggestions pointed to by users in our study; and to expose more elements of the UE Editor to the system.

\paragraph{Direct manipulation in the editor}

Though the system can currently observe edits to code made by users, it could also conceivably respond to modifications made to the placement and other features of objects within the editor.
Currently, the system generates the initial layout (in terms of position coordinates, scale, and rotation) and initial parameters (e.g. the initial speed or direction of a moving object) of actor instances by producing a structured json containing this information.
Functionality could be added to record changes made to these parameters, or the addition/deletion of actors, by the user to the level during their interaction with the editor, such that the next iteration of code generation would take this updated layout information as input.

We note a general frustration with the inability to directly observe generated artifacts in the UE editor.
This is largely an engineering challenge that could conceivably be overcome in future iterations using one of several approaches.
Most notably, we are currently not compiling in UE's ``LiveCode'' mode from our python backend, though this is possible in theory (instead, we are forced to compile with the UE Editor closed, then relaunch it each time new code is generated).
This change would allow the editor to remain open, even as our system was in the process or editing C++ Actor classes or adding new textured meshes to the scene.
This could allow for a more fluid ``dialogue'' between human users and the system,  with the user potentially making changes to generated code or layouts inside the Editor, then having the system iterate on the project while taking these user edits into account.

In addition or separately from seeing newly generated artifacts reflected in the Editor, the user may also want the option of exploring partially completed environments while at the same time letting the system ``work ahead''. 
Currently, the option to ``compile and run'' partially-complete environments, presented as a button on visual feedback nodes in the GUI, requires pausing the backend, since collecting visual feedback at future steps ends with killing UE processes, which would result in killing the instance of the engine intended for the user.
This could likely be addressed by tracking which process ID belongs to different instances of the Editor, and killing only the appropriate processes from the backend.


\paragraph{GUI improvements}
Uncertainty about node colors demonstrates the need for a legend, or for more explicit node type labelling.
Further, colors like green, yellow and red should likely be reserved to indicate (partial) success/failure of artifacts generated by the coding submodule, which is able to iterate on errors in its output.
The system's operation could also easily be made more transparent by highlighting the node currently in-progress.

\paragraph{Ethics}
\label{sec:ethics}
The use of LLMs that power DreamGarden raises important ethical concerns. These models are trained on vast datasets, often collected without explicit consent, raising questions about data ownership. This data can contain biases perpetuated by models leading to harmful or inaccurate outputs. 
Finally, significant computational resources are required to train these models contributing considerably to carbon emissions. To address these challenges, DreamGarden incorporates user feedback loops, giving designers control to intervene and guide the system away from unintended results. We intentionally avoid fine-tuning models, focusing on prompting existing models which can cut energy and computing use by at least 1000 times \cite{Martineau_2023}. This allowed us to keep the API costs during development and user study to under \$200. The system is also designed to be modular, allowing models to be swapped out as more energy-efficient ones become available, ensuring that it evolves alongside advancements in sustainable AI.



\section{Conclusion \& Future Work}

\textit{DreamGarden} demonstrates the promise of interfacing multi-agent LLM-driven systems with game engines for open-ended content creation.
The modular nature of generated plan and action trees opens the door to yet more forms of user interaction beyond prompting, pruning and expansion.
Future work could generate (or solicit from users) multi-modal sketches (e.g. rough images of levels, diagrams of mechanics, or toy 2D prototype games) which could be used to guide the later generation of the final experience in Unreal Engine, or incorporate additional auxiliary tools (such as Blender~\cite{blender} or Houdini~\cite{houdini}) for specialized content generation via implementation submodules.

Meanwhile, the recursive growth of a tree of plans to be implemented by predefined submodules is a promising and general approach to multi-agent orchestration in complex domains. Future work could investigate fine-tuning LLMs on proven high level-implementation plans of existing projects or domain-specific knowledge, both in the gaming industry and in software development more broadly.

Regarding our initial research questions, we find that, while the system often struggles to autonomously generate complex simulations (Q1), users find substantial value in the intermediary planning steps and artifacts it generates in the process (Q2).
Despite user studies being limited to a relatively small amount of time, we note the emergence of diverse modes of user interaction with the system (Q3).
Inspired by the experiences and feedback from users in our usability study, we also conceptualized various means of fleshing out these modes of interaction to bolster DreamGarden's co-creative potential.

In particular, we note the strong appeal of such a branching approach to ``growing'' content collaboratively with AI agents, and argue that a graph-based approach is a natural fit for creative and open-ended tasks.
We also see that, even given our relatively rudimentary GUI, the paradigm of ``gardening'' as opposed to constant active human guidance or a fully passive ``prompt and sit back'' approach, invites a variety of modes of interaction from users with differing intents and levels of experience.
The key here is in combining autonomous feedback and self-improvement with a variety of means of user intervention.
This is strengthened by our focus on having the system operate on human-interpretable representations of generated content inside an established game engine, in contrast to extant efforts that seek to replace game engines with, effectively, uninterpretable latent variables.

Our work here suggests clear directions for future work, chief among them increasing the means of interaction with such agentic systems to go beyond reliance on simple prompts and keeping the user more engaged and informed of the intermediary implementation steps. One can imagine a much more general mode of interaction on growing computational graphs and parallelization of independent tasks. Of particular interest for future work would be a generalization of  \textit{DreamGarden} that can build on existing Unreal projects with human-built complex game mechanics, narratives and all the other elements that make up a game to streamline creation of additional content, e.g. new quests in a role-playing game, or adapting existing content to a new gameplay mechanic. Future versions of \textit{DreamGarden} could also be built with a much narrower scope that focuses on concrete core game design processes, rather than attempting a full end-to-end system.
Finally, a more fully autonomous version of \textit{DreamGarden} could be developed, in which an LLM-driven meta-planner generates and iterates on the seed prompts themselves, pre-populating a pool of novel simulations which human designers could use as starting points.


\begin{acks}
We would like to thank various collaborators who gave very insightful comments and suggestions, in particular Jaron Lanier, Jianfeng Gao, Julian Togelius, Judith Amores, Amina Kobenova, Cyan DeVeaux, Jennifer Marsman, Haiyan Zhang, Nebojsa Jojic, Andy Wilson. Additionally, we would like to thank all the study participants. 
\end{acks}

\bibliographystyle{ACM-Reference-Format}
\bibliography{ref}




\appendix

\section{Walkthrough example prompts}

\begin{lstlisting}[label={lst:planner_example_prompt},caption={Example prompts for broad plan generation}]
SYSTEM PROMPT:
You are a game design assistant. The user will describe a dream, an imagined scenario or game, a memory, etc., and your job is to reason about how this base text can be turned into a video game. In this initial planning phase, you will generate a high-level plan (with no more than 3 steps) for how to turn the base text into a game. (Later, this plan will be broken down into sub-steps.) First describe the game and conceptualize the overall design. Then, create a high-level, multi-step plan for the development process.For now, let's focus on generating a simple sketch version of this game. It must be a limited to a zero-player simulation, which makes sense when viewed from a single camera angle in Unreal Engine. Note that later, in the hierarchical game design plan tree, each leaf node will ultimately correspond to the application of one of the following submodules:
- ASSET DOWNLOAD: This module downloads and imports a textured mesh from objaverse, an online asset library.
- ACTOR GENERATION: This module writes C++ code for one or several actor classes in Unreal Engine, and generates a structured representation of their initial spawning layout and parameters in the scene.
- PROCEDURAL MESH GENERATION: This module writes code to procedurally generate an asset consisting of a textured mesh in UE.
- DIFFUSION MESH GENERATION: This module produces a text description of an asset, which is turned into a textured 3D mesh using a generative (diffusion) model. In each leaf node, it can generate only ONE object. It should only be used to generate meshes for entities that are singular functional components. E.g., if a character has a sword which can be separated from the character during gameplay, the sword should be generated separately from the character.)
So don't come up with steps for doing something that cannot ultimately carried about by these submodules (e.g. marketing or sales). (At the same time, the steps of your high-level plan should not invoke these submodules directly). Your response should follow the following formatting, e.g.:

Game Conceptualization:
The game involves capturing rare creatures with special abilities and having them compete against one another in battles. The aesthetic will be... etc.

Game Design Plan:
1. Design overworld layout
2. Implement game mechanics
3. Design character assets...etc.

Later, a separate agent will expand this tree to involve more fine-grained implementation steps. 

USER PROMPT:
Generate a high-level game development plan for the following base text: A spaceship flies around the sky, abducting cows and other animals. The cows are inside an enclosure behind a farm.
\end{lstlisting}

\begin{lstlisting}[label={lst:subplanner_example_prompt},caption={Example prompts for (one step of) recursive sub-plan generation}]
SYSTEM PROMPT:
You are a game design assistant. The user will describe a dream, an imagined scenario or game, a memory, etc., and your job is to reason about how this base text can be turned into a video game. In this phase of planning, you are given a high-level plan, and will be asked to break each step of the plan down into sub-steps, recursively, resulting in a tree where each leaf node involves the generation of some asset or code (C++ for Unreal Engine), and such that all nodes executed in sequence will produce the code and assets for a functional simulation. You will be asked to produce the children for a node, one node at a time. You will be given the entire plan/tree in its current state, and the node marked for elaboration will be specified. First, reason about what the node entails (taking into account the broader context of the game), then break it into child nodes. If a child node is a leaf (i.e. involves some concrete and specific asset or code generation task), mark it as such, and assign it to the relevant submodule. Your response should follow the following formatting, e.g.,

Node Description: Implement zombie NPC
Sub-Steps:
1. Generate zombie NPC asset. LEAF. DIFFUSION MESH GENERATION.
2.Implement zombie NPC pathfinding, enemy detection, and random walking behavior.
3. ...

NOTE: For now, let's focus on generating a simple sketch version of this game. It must be a limited to a zero-player simulation, which makes sense when viewed from a single camera angle in Unreal Engine. In the hierarchical game design plan tree, each leaf node will ultimately correspond to the application of one of the following submodules:
- ASSET DOWNLOAD: This module downloads and imports a textured mesh from objaverse, an online asset library.
- ACTOR GENERATION: This module writes C++ code for one or several actor classes in Unreal Engine, and generates a structured representation of their initial spawning layout and parameters in the scene.
- PROCEDURAL MESH GENERATION: This module writes code to procedurally generate an asset consisting of a textured mesh in UE.
- DIFFUSION MESH GENERATION: This module produces a text description of an asset, which is turned into a textured 3D mesh using a generative (diffusion) model. In each leaf node, it can generate only ONE object. It should only be used to generate meshes for entities that are singular functional components. E.g., if a character has a sword which can be separated from the character during gameplay, the sword should be generated separately from the character.

Be sure to generate any assets before implementing actors for these assets, so that the actors can correctly reference the generated meshes.Mark each leaf node with the name of the submodule to which it corresponds.

IMPORTANT: The tree should not have a branching factor greater than 3 nor depth greater than 3. So any list you output much be limited to 3 elements, and any list item you generate that has 2 parents must be marked a LEAF node.

USER PROMPT:
The current game development plan is as follows:
The game is a zero-player simulation where a spaceship flies around the sky, periodically descending to abduct cows and other animals from an enclosure behind a farm. The aesthetic will be whimsical and light-hearted, with cartoonish graphics and humorous animations. The camera will be positioned to provide a clear view of the farm, enclosure, and spaceship's movements. The primary focus will be on the interactions between the spaceship and the animals, creating an entertaining spectacle for the viewer.

1.  Design the environment layout
  1.1.  Generate the farm environment asset. LEAF. DIFFUSION MESH GENERATION.
  1.2.  Generate the animal enclosure asset. LEAF. DIFFUSION MESH GENERATION.
  1.3.  Arrange the environment in the scene. LEAF. ACTOR GENERATION.
2.  Implement spaceship behavior
3.  Design animal assets and their interactions


Please elaborate on the following node, producing a depth-1 list of sub-nodes. DO NOT PROVIDE ANY SUB-STEPS BEYOND THIS FIRST LAYER OF CHILDREN! Current node:
2:  Implement spaceship behavior
\end{lstlisting}

\begin{lstlisting}[label={lst:task_generator_example_prompt_code},caption={Example prompt for the task generator (tasked in this case with the production of a prompt to be passed to the code generation submodule.}]
SYSTEM PROMPT:
You are a game development assistant. You will be presented with a sequence of steps required to implement a game. Your job will be to take a given step in the list, and rephrase it as a detailed and specific prompt for a game development submodule.

USER PROMPT:
The list of implementation prompts is as follows:
ASSET_NAME: FarmEnvironment
ASSET_DESCRIPTION: Generate an image of a detailed farm environment asset with a variety of elements. Include a large barn with a red exterior and white trim, a wooden fence enclosing a green pasture, a silo, a dirt path leading to the barn, and a water trough. Ensure the scene is well-lit to highlight textures and details. The image should be without a background, high resolution, and suitable for use in Unreal Engine.
ASSET_NAME: AnimalEnclosure
ASSET_DESCRIPTION: Generate an image of an animal enclosure asset with detailed features. Include a sturdy wooden fence with vertical planks and horizontal support beams, a secure gate with a latch, and a small wooden shelter within the enclosure. The fence should enclose a grassy area with slight variations in terrain. Ensure the scene is well-lit to highlight textures and details. The image should be without a background, high resolution, and suitable for use in Unreal Engine.

 The remaining implementation steps are:
2.  Arrange the environment in the scene. LEAF. ACTOR GENERATION.
3.  Generate spaceship asset. LEAF. DIFFUSION MESH GENERATION.
4.  Implement spaceship initial flight path and movement logic. LEAF. ACTOR GENERATION.
5.  Implement periodic descent behavior for the spaceship. LEAF. ACTOR GENERATION.
6.  Implement randomized flight patterns and altitude changes. LEAF. ACTOR GENERATION.
7.  Implement logic for detecting animals within the spaceship's abduction range. LEAF. ACTOR GENERATION.
8.  Implement animation and effects for the abduction process (e.g., beam effect). LEAF. ACTOR GENERATION.
9.  Implement behavior for removing the abducted animals from the scene. LEAF. ACTOR GENERATION.
10.  Generate cow asset. LEAF. DIFFUSION MESH GENERATION.
11.  Generate other animal assets (e.g., sheep, pigs, chickens). LEAF. DIFFUSION MESH GENERATION.
12.  Implement animal behavior and animations. LEAF. ACTOR GENERATION. 

Write an implementation prompt for step 2:  Arrange the environment in the scene. LEAF. ACTOR GENERATION..
You are prompting a submodule that writes C++ code for actors in Unreal Engine 5, and lays them out in a scene. For the first item in the list, the coding agent will be creating new files, while for following items, the coder might be creating new files or editing existing ones. Crucially, we want to formulate each implementation step such that all the code taken together should be compilable and testable in Unreal Engine. This may mean that your final prompts may involve certain placeholder elements or functionalities to be fleshed out at later steps.  Format your response exactly as follows:
CODE_PROMPT: Write/extend an actor class that...
SPAWN_PROMPT: Place the actor in the scene at...
\end{lstlisting}

\begin{lstlisting}[label={lst:task_generator_example_prompt},caption={Example prompt for the task generator (tasked in this case with the prodution of a prompt to be passed to the 2D-to-3D diffusion mesh generator).}]
SYSTEM PROMPT:
You are a game development assistant. You will be presented with a sequence of steps required to implement a game. Your job will be to take a given step in the list, and rephrase it as a detailed and specific prompt for a game development submodule.

USER PROMPT:
The list of implementation prompts is as follows:
ASSET_NAME: FarmEnvironment
ASSET_DESCRIPTION: Generate an image of a detailed farm environment asset with a variety of elements. Include a large barn with a red exterior and white trim, a wooden fence enclosing a green pasture, a silo, a dirt path leading to the barn, and a water trough. Ensure the scene is well-lit to highlight textures and details. The image should be without a background, high resolution, and suitable for use in Unreal Engine.
ASSET_NAME: AnimalEnclosure
ASSET_DESCRIPTION: Generate an image of an animal enclosure asset with detailed features. Include a sturdy wooden fence with vertical planks and horizontal support beams, a secure gate with a latch, and a small wooden shelter within the enclosure. The fence should enclose a grassy area with slight variations in terrain. Ensure the scene is well-lit to highlight textures and details. The image should be without a background, high resolution, and suitable for use in Unreal Engine.
CODE_PROMPT: Write an actor class that represents the farm environment, including the large barn with a red exterior and white trim, a wooden fence enclosing a green pasture, a silo, a dirt path leading to the barn, and a water trough. Ensure that the class includes properties for each of these elements and that they are properly instantiated and positioned relative to each other to form a cohesive environment. Use appropriate Unreal Engine components such as StaticMeshComponent for the barn, fence, silo, and trough, and ensure the textures and materials are correctly applied. The actor class should also include functionality to ensure the scene is well-lit to highlight textures and details.
SPAWN_PROMPT: Place the farm environment actor in the scene at coordinates (X=0, Y=0, Z=0) ensuring it is centered and properly aligned with the ground plane.

 The remaining implementation steps are:
3.  Generate spaceship asset. LEAF. DIFFUSION MESH GENERATION.
4.  Implement spaceship initial flight path and movement logic. LEAF. ACTOR GENERATION.
5.  Implement periodic descent behavior for the spaceship. LEAF. ACTOR GENERATION.
6.  Implement randomized flight patterns and altitude changes. LEAF. ACTOR GENERATION.
7.  Implement logic for detecting animals within the spaceship's abduction range. LEAF. ACTOR GENERATION.
8.  Implement animation and effects for the abduction process (e.g., beam effect). LEAF. ACTOR GENERATION.
9.  Implement behavior for removing the abducted animals from the scene. LEAF. ACTOR GENERATION.
10.  Generate cow asset. LEAF. DIFFUSION MESH GENERATION.
11.  Generate other animal assets (e.g., sheep, pigs, chickens). LEAF. DIFFUSION MESH GENERATION.
12.  Implement animal behavior and animations. LEAF. ACTOR GENERATION. 

Write an implementation prompt for step 3:  Generate spaceship asset. LEAF. DIFFUSION MESH GENERATION..
You are generating a text prompt for an asset for generating submodules. First, your prompt will be turned into several 2D images. The most promising of these will be selected by a separate agent or human user, then fed to a 2D-to-3D generative model to produce a textured mesh, which will then be imported in Unreal Engine. Therefore, you should prompt the 2D generative model in such a way that it generates images which completely and clearly display the relevant features of a 3D object, such that it can be reasonably represented by the 2D-to-3D model. Generate only ONE object. Prompt for an image with no background, and a high resolution of the type which would look plausible in Unreal Engine. Format the prompt exactly as follows:
ASSET NAME: LargeBoulder
IMAGE PROMPT: Generate an image of a large boulder...
\end{lstlisting}

\begin{lstlisting}[caption={Example prompt for the code generator. Note that this includes paths to generated assets, the latest scene layout, and all code generated thus far in the sequential execution of implementation tasks.}]
SYSTEM_PROMPT:
You are a game development agent, capable of writing complete and functional C++ code for Unreal Engine 5. You will be asked to implement a particular feature. The project name is a00. Write all code necessary such that when the project is compiled the feature is fully functional. The target feature should be visible in the simulation given your code alone. Some notes:
- When writing actors with editable properties, you must set reasonable default values. DO NOT assume that a human will manually manipulate the editor before the game is run.
- Include ample logging and warnings so that, when an LLM agent evaluates the Unreal Engine log, it can make useful conclusions about possible issues in the code. For example, if a mesh or material is not loaded successfully, `UE_LOG` warning should be printed.
- Ensure that `FObjectFinders` are only used inside constructors.
- Ultimately, the correctness of your code will be evaluated by visual feedback. So, be sure to use visually obvious placeholder assets if no appropriate ones available. Also be sure that all relevant mechanics take place within the first 6 seconds of simulation. Assets and timing can always be adjusted later
 - You cannot edit the `Build.cs` file, so if, in prior code, you tried to import a third-party plugin and ran into an error, you should find an alternative (i.e. implement the desired functionality from scratch).
 - You are not capable of animating or importing skeletal meshes (moving static meshes around is enough for now)
- Some default lighting is already present in the scene.

Some code has already been written, and your job will be to extend it to implement a new target feature. You may output either complete files or `diff' edits. When outputting entire files, follow exactly this formatting to indicate filenames and code blocks (for parsing via regex):

 FILENAME: MyFile.cpp

[CODE HERE]

FILENAME: MyFile.h

[CODE HERE]...
When outputting `diff' edits, follow this format:
```diff
- old code line 1
- old code line 2
+ new code line 1
+ new code line 2
```

The following materials are available in `/Game/StarterContent/Materials`: M_AssetPlatform.uasset, M_Basic_Floor.uasset, M_Basic_Wall.uasset, M_Brick_Clay_Beveled.uasset, M_Brick_Clay_New.uasset, M_Brick_Clay_Old.uasset, M_Brick_Cut_Stone.uasset, M_Brick_Hewn_Stone.uasset, M_Ceramic_Tile_Checker.uasset, M_CobbleStone_Pebble.uasset, M_CobbleStone_Rough.uasset, M_CobbleStone_Smooth.uasset, M_ColorGrid_LowSpec.uasset, M_Concrete_Grime.uasset, M_Concrete_Panels.uasset, M_Concrete_Poured.uasset, M_Concrete_Tiles.uasset, M_Glass.uasset, M_Ground_Grass.uasset, M_Ground_Gravel.uasset, M_Ground_Moss.uasset, M_Metal_Brushed_Nickel.uasset, M_Metal_Burnished_Steel.uasset, M_Metal_Chrome.uasset, M_Metal_Copper.uasset, M_Metal_Gold.uasset, M_Metal_Rust.uasset, M_Metal_Steel.uasset, M_Rock_Basalt.uasset, M_Rock_Marble_Polished.uasset, M_Rock_Sandstone.uasset, M_Rock_Slate.uasset, M_Tech_Checker_Dot.uasset, M_Tech_Hex_Tile.uasset, M_Tech_Hex_Tile_Pulse.uasset, M_Tech_Panel.uasset, M_Water_Lake.uasset, M_Water_Ocean.uasset, M_Wood_Floor_Walnut_Polished.uasset, M_Wood_Floor_Walnut_Worn.uasset, M_Wood_Oak.uasset, M_Wood_Pine.uasset, M_Wood_Walnut.uasset
These are the ONLY materials you may use in the scene.
Here is an example of how to correctly load a material from the StarterContent folder:
`ConstructorHelpers::FObjectFinder<UMaterialInterface> MaterialAsset(TEXT("/Game/StarterContent/Materials/M_Ground_Grass.M_Ground_Grass"));`


The following meshes are available in `/Game/StarterContent/Props`: MaterialSphere.uasset, SM_Bush.uasset, SM_Chair.uasset, SM_CornerFrame.uasset, SM_Couch.uasset, SM_Door.uasset, SM_DoorFrame.uasset, SM_GlassWindow.uasset, SM_Lamp_Ceiling.uasset, SM_Lamp_Wall.uasset, SM_PillarFrame.uasset, SM_PillarFrame300.uasset, SM_Rock.uasset, SM_Shelf.uasset, SM_Stairs.uasset, SM_Statue.uasset, SM_TableRound.uasset, SM_WindowFrame.uasset
The following meshes are available in `/Game/StarterContent/Shapes`: Shape_Cone.uasset, Shape_Cube.uasset, Shape_Cylinder.uasset, Shape_NarrowCapsule.uasset, Shape_Pipe.uasset, Shape_Pipe_180.uasset, Shape_Pipe_90.uasset, Shape_Plane.uasset, Shape_QuadPyramid.uasset, Shape_Sphere.uasset, Shape_Torus.uasset, Shape_Trim.uasset, Shape_Trim_90_In.uasset, Shape_Trim_90_Out.uasset, Shape_TriPyramid.uasset, Shape_Tube.uasset, Shape_Wedge_A.uasset, Shape_Wedge_B.uasset, Shape_WideCapsule.uasset
The following meshes are available in `/Game/StarterContent/Architecture`: Floor_400x400.uasset, Pillar_50x500.uasset, SM_AssetPlatform.uasset, Wall_400x200.uasset, Wall_400x300.uasset, Wall_400x400.uasset, Wall_500x500.uasset, Wall_Door_400x300.uasset, Wall_Door_400x400.uasset, Wall_Window_400x300.uasset, Wall_Window_400x400.uasset
The following meshes are available in `/Game/GeneratedMeshes`: AnimalEnclosure.uasset, FarmEnvironment.uasset, Sheep.uasset, Spaceship.uasset
These are the ONLY meshes you may use in the scene.
Here is an example of how to correctly load a mesh from the StarterContent folder:
`static ConstructorHelpers::FObjectFinder<UStaticMesh> Mesh(TEXT("/Game/StarterContent/Shapes/Shape_Sphere.Shape_Sphere"));`

USER_PROMPT:
Our task is as follows:
```Write an actor class that represents the animal behavior and animations. Create a base class named `AnimalActor` that includes properties for common animal characteristics, such as health, movement speed, and animations. Implement methods for basic behaviors such as walking, eating, and idling. Use Unreal Engine's Animation Blueprint to handle the animation states and transitions. Ensure the class includes methods to trigger different animations based on the animal's state and behavior. Use Unreal Engine components such as CharacterMovementComponent for movement and SkeletalMeshComponent for the animal mesh and animations. Ensure the animations are frame-rate independent by using DeltaTime in your calculations.
```
Here is the existing code:

FILENAME: FarmEnvironmentActor.h
```cpp
#pragma once

#include "CoreMinimal.h"
#include "GameFramework/Actor.h"
#include "FarmEnvironmentActor.generated.h"

UCLASS()
class A00_API AFarmEnvironmentActor : public AActor
{
    GENERATED_BODY()

public:
    AFarmEnvironmentActor();

protected:
    virtual void BeginPlay() override;

public:
    virtual void Tick(float DeltaTime) override;

private:
    UPROPERTY(VisibleAnywhere)
    UStaticMeshComponent* BarnMesh;

    UPROPERTY(VisibleAnywhere)
    UStaticMeshComponent* FenceMesh;

    UPROPERTY(VisibleAnywhere)
    UStaticMeshComponent* SiloMesh;

    UPROPERTY(VisibleAnywhere)
    UStaticMeshComponent* PathMesh;

    UPROPERTY(VisibleAnywhere)
    UStaticMeshComponent* TroughMesh;

    UPROPERTY(VisibleAnywhere)
    UMaterialInterface* BarnMaterial;

    UPROPERTY(VisibleAnywhere)
    UMaterialInterface* FenceMaterial;

    UPROPERTY(VisibleAnywhere)
    UMaterialInterface* SiloMaterial;

    UPROPERTY(VisibleAnywhere)
    UMaterialInterface* PathMaterial;

    UPROPERTY(VisibleAnywhere)
    UMaterialInterface* TroughMaterial;

    UPROPERTY(VisibleAnywhere)
    UMaterialInterface* PastureMaterial;
};
```
FILENAME: FarmEnvironmentActor.cpp
```cpp
#include "FarmEnvironmentActor.h"
#include "Components/StaticMeshComponent.h"
#include "UObject/ConstructorHelpers.h"

// Sets default values
AFarmEnvironmentActor::AFarmEnvironmentActor()
{
    PrimaryActorTick.bCanEverTick = true;

    // Initialize Barn
    BarnMesh = CreateDefaultSubobject<UStaticMeshComponent>(TEXT("BarnMesh"));
    static ConstructorHelpers::FObjectFinder<UStaticMesh> BarnMeshAsset(TEXT("/Game/StarterContent/Architecture/Wall_500x500.Wall_500x500"));
    if (BarnMeshAsset.Succeeded())
    {
        BarnMesh->SetStaticMesh(BarnMeshAsset.Object);
        BarnMesh->SetupAttachment(RootComponent);
        BarnMesh->SetRelativeLocation(FVector(0.0f, 0.0f, 0.0f));
    }
    static ConstructorHelpers::FObjectFinder<UMaterialInterface> BarnMaterialAsset(TEXT("/Game/StarterContent/Materials/M_Brick_Clay_Old.M_Brick_Clay_Old"));
    if (BarnMaterialAsset.Succeeded())
    {
        BarnMaterial = BarnMaterialAsset.Object;
        BarnMesh->SetMaterial(0, BarnMaterial);
    }

    // Initialize Fence
    FenceMesh = CreateDefaultSubobject<UStaticMeshComponent>(TEXT("FenceMesh"));
    static ConstructorHelpers::FObjectFinder<UStaticMesh> FenceMeshAsset(TEXT("/Game/StarterContent/Architecture/Wall_400x200.Wall_400x200"));
    if (FenceMeshAsset.Succeeded())
    {
        FenceMesh->SetStaticMesh(FenceMeshAsset.Object);
        FenceMesh->SetupAttachment(RootComponent);
        FenceMesh->SetRelativeLocation(FVector(500.0f, 0.0f, 0.0f));
        FenceMesh->SetRelativeScale3D(FVector(0.25f, 1.0f, 0.5f));
    }
    static ConstructorHelpers::FObjectFinder<UMaterialInterface> FenceMaterialAsset(TEXT("/Game/StarterContent/Materials/M_Wood_Pine.M_Wood_Pine"));
    if (FenceMaterialAsset.Succeeded())
    {
        FenceMaterial = FenceMaterialAsset.Object;
        FenceMesh->SetMaterial(0, FenceMaterial);
    }

    // Initialize Silo
    SiloMesh = CreateDefaultSubobject<UStaticMeshComponent>(TEXT("SiloMesh"));
    static ConstructorHelpers::FObjectFinder<UStaticMesh> SiloMeshAsset(TEXT("/Game/StarterContent/Shapes/Shape_Cylinder.Shape_Cylinder"));
    if (SiloMeshAsset.Succeeded())
    {
        SiloMesh->SetStaticMesh(SiloMeshAsset.Object);
        SiloMesh->SetupAttachment(RootComponent);
        SiloMesh->SetRelativeLocation(FVector(1000.0f, 0.0f, 0.0f));
        SiloMesh->SetRelativeScale3D(FVector(1.0f, 1.0f, 3.0f));
    }
    static ConstructorHelpers::FObjectFinder<UMaterialInterface> SiloMaterialAsset(TEXT("/Game/StarterContent/Materials/M_Metal_Steel.M_Metal_Steel"));
    if (SiloMaterialAsset.Succeeded())
    {
        SiloMaterial = SiloMaterialAsset.Object;
        SiloMesh->SetMaterial(0, SiloMaterial);
    }

    // Initialize Path
    PathMesh = CreateDefaultSubobject<UStaticMeshComponent>(TEXT("PathMesh"));
    static ConstructorHelpers::FObjectFinder<UStaticMesh> PathMeshAsset(TEXT("/Game/StarterContent/Shapes/Shape_Plane.Shape_Plane"));
    if (PathMeshAsset.Succeeded())
    {
        PathMesh->SetStaticMesh(PathMeshAsset.Object);
        PathMesh->SetupAttachment(RootComponent);
        PathMesh->SetRelativeLocation(FVector(0.0f, -250.0f, -50.0f));
        PathMesh->SetRelativeScale3D(FVector(10.0f, 1.0f, 1.0f));
    }
    static ConstructorHelpers::FObjectFinder<UMaterialInterface> PathMaterialAsset(TEXT("/Game/StarterContent/Materials/M_Concrete_Grime.M_Concrete_Grime"));
    if (PathMaterialAsset.Succeeded())
    {
        PathMaterial = PathMaterialAsset.Object;
        PathMesh->SetMaterial(0, PathMaterial);
    }

    // Initialize Trough
    TroughMesh = CreateDefaultSubobject<UStaticMeshComponent>(TEXT("TroughMesh"));
    static ConstructorHelpers::FObjectFinder<UStaticMesh> TroughMeshAsset(TEXT("/Game/StarterContent/Shapes/Shape_Cube.Shape_Cube"));
    if (TroughMeshAsset.Succeeded())
    {
        TroughMesh->SetStaticMesh(TroughMeshAsset.Object);
        TroughMesh->SetupAttachment(RootComponent);
        TroughMesh->SetRelativeLocation(FVector(0.0f, 500.0f, -25.0f));
        TroughMesh->SetRelativeScale3D(FVector(2.0f, 0.5f, 0.25f));
    }
    static ConstructorHelpers::FObjectFinder<UMaterialInterface> TroughMaterialAsset(TEXT("/Game/StarterContent/Materials/M_Metal_Rust.M_Metal_Rust"));
    if (TroughMaterialAsset.Succeeded())
    {
        TroughMaterial = TroughMaterialAsset.Object;
        TroughMesh->SetMaterial(0, TroughMaterial);
    }

    // Initialize Pasture
    UStaticMeshComponent* PastureMesh = CreateDefaultSubobject<UStaticMeshComponent>(TEXT("PastureMesh"));
    static ConstructorHelpers::FObjectFinder<UStaticMesh> PastureMeshAsset(TEXT("/Game/StarterContent/Shapes/Shape_Plane.Shape_Plane"));
    if (PastureMeshAsset.Succeeded())
    {
        PastureMesh->SetStaticMesh(PastureMeshAsset.Object);
        PastureMesh->SetupAttachment(RootComponent);
        PastureMesh->SetRelativeLocation(FVector(0.0f, 0.0f, -50.0f));
        PastureMesh->SetRelativeScale3D(FVector(20.0f, 20.0f, 1.0f));
    }
    static ConstructorHelpers::FObjectFinder<UMaterialInterface> PastureMaterialAsset(TEXT("/Game/StarterContent/Materials/M_Ground_Grass.M_Ground_Grass"));
    if (PastureMaterialAsset.Succeeded())
    {
        PastureMaterial = PastureMaterialAsset.Object;
        PastureMesh->SetMaterial(0, PastureMaterial);
    }

    // Set Root Component
    RootComponent = PastureMesh;
}

// Called when the game starts or when spawned
void AFarmEnvironmentActor::BeginPlay()
{
    Super::BeginPlay();
}

// Called every frame
void AFarmEnvironmentActor::Tick(float DeltaTime)
{
    Super::Tick(DeltaTime);
}
```
FILENAME: SpaceshipActor.h
```cpp
#pragma once

#include "CoreMinimal.h"
#include "GameFramework/Actor.h"
#include "Particles/ParticleSystemComponent.h"
#include "Components/PointLightComponent.h"
#include "Components/StaticMeshComponent.h"
#include "SpaceshipActor.generated.h"

UCLASS()
class A00_API ASpaceshipActor : public AActor
{
    GENERATED_BODY()

public:
    ASpaceshipActor();

protected:
    virtual void BeginPlay() override;

public:
    virtual void Tick(float DeltaTime) override;

    void SetInitialFlightPath(FVector StartLocation, FVector TargetLocation);

private:
    void StartDescent();
    void TriggerDescent();
    void StartRandomMovement();
    void ChangeFlightPattern();
    void ChangeAltitude();
    void DetectAnimals();
    void OnDetectAnimals();
    void StartAbductionProcess(AActor* Animal);
    void AnimateAbduction(float DeltaTime);
    void RemoveAbductedAnimal();

    UPROPERTY(VisibleAnywhere)
    UStaticMeshComponent* SpaceshipMesh;

    UPROPERTY(EditAnywhere, BlueprintReadWrite, Category="Flight", meta=(AllowPrivateAccess="true"))
    float Speed;

    UPROPERTY(EditAnywhere, BlueprintReadWrite, Category="Flight", meta=(AllowPrivateAccess="true"))
    float SpeedVariance;

    UPROPERTY(EditAnywhere, BlueprintReadWrite, Category="Flight", meta=(AllowPrivateAccess="true"))
    float DirectionVariance;

    UPROPERTY(EditAnywhere, BlueprintReadWrite, Category="Descent", meta=(AllowPrivateAccess="true"))
    float DescentFrequency;

    UPROPERTY(EditAnywhere, BlueprintReadWrite, Category="Descent", meta=(AllowPrivateAccess="true"))
    float DescentSpeed;

    UPROPERTY(EditAnywhere, BlueprintReadWrite, Category="Descent", meta=(AllowPrivateAccess="true"))
    float DescentDistance;

    UPROPERTY(EditAnywhere, BlueprintReadWrite, Category="Flight", meta=(AllowPrivateAccess="true"))
    float AltitudeChangeFrequency;

    UPROPERTY(EditAnywhere, BlueprintReadWrite, Category="Flight", meta=(AllowPrivateAccess="true"))
    float MinAltitude;

    UPROPERTY(EditAnywhere, BlueprintReadWrite, Category="Flight", meta=(AllowPrivateAccess="true"))
    float MaxAltitude;

    UPROPERTY(EditAnywhere, BlueprintReadWrite, Category="Detection", meta=(AllowPrivateAccess="true"))
    float AbductionRange;

    UPROPERTY(VisibleAnywhere, Category="Detection")
    TArray<AActor*> DetectedAnimals;

    UPROPERTY(VisibleAnywhere, Category="Abduction")
    UParticleSystemComponent* AbductionBeam;

    UPROPERTY(VisibleAnywhere, Category="Abduction")
    UPointLightComponent* BeamLight;

    FVector Direction;
    FVector TargetLocation;
    bool bIsFlying;
    bool bIsDescending;
    FVector DescentTargetLocation;

    FTimerHandle DescentTimerHandle;
    FTimerHandle FlightPatternTimerHandle;
    FTimerHandle AltitudeChangeTimerHandle;
    FTimerHandle DetectionTimerHandle;

    AActor* AbductingAnimal;
    bool bIsAbducting;
};
```
FILENAME: SpaceshipActor.cpp
```cpp
#include "SpaceshipActor.h"
#include "UObject/ConstructorHelpers.h"
#include "TimerManager.h"
#include "Engine/World.h"
#include "Engine/Engine.h"
#include "DrawDebugHelpers.h"
#include "Math/UnrealMathUtility.h"
#include "Engine/OverlapResult.h"
#include "Components/SphereComponent.h"
#include "GameFramework/Actor.h"
#include "Particles/ParticleSystem.h"

// Sets default values
ASpaceshipActor::ASpaceshipActor()
{
    PrimaryActorTick.bCanEverTick = true;

    // Initialize Spaceship Mesh
    SpaceshipMesh = CreateDefaultSubobject<UStaticMeshComponent>(TEXT("SpaceshipMesh"));
    static ConstructorHelpers::FObjectFinder<UStaticMesh> SpaceshipMeshAsset(TEXT("/Game/GeneratedMeshes/Spaceship.Spaceship"));
    if (SpaceshipMeshAsset.Succeeded())
    {
        SpaceshipMesh->SetStaticMesh(SpaceshipMeshAsset.Object);
        RootComponent = SpaceshipMesh;
    }
    else
    {
        UE_LOG(LogTemp, Warning, TEXT("Spaceship mesh not found!"));
    }

    // Initialize Abduction Beam
    AbductionBeam = CreateDefaultSubobject<UParticleSystemComponent>(TEXT("AbductionBeam"));
    static ConstructorHelpers::FObjectFinder<UParticleSystem> AbductionBeamAsset(TEXT("/Game/StarterContent/Particles/P_Fire.P_Fire"));
    if (AbductionBeamAsset.Succeeded())
    {
        AbductionBeam->SetTemplate(AbductionBeamAsset.Object);
        AbductionBeam->SetupAttachment(RootComponent);
        AbductionBeam->SetRelativeLocation(FVector(0.0f, 0.0f, -100.0f));
        AbductionBeam->SetVisibility(false);
    }
    else
    {
        UE_LOG(LogTemp, Warning, TEXT("Abduction beam particle system not found!"));
    }

    // Initialize Beam Light
    BeamLight = CreateDefaultSubobject<UPointLightComponent>(TEXT("BeamLight"));
    BeamLight->SetupAttachment(RootComponent);
    BeamLight->SetRelativeLocation(FVector(0.0f, 0.0f, -100.0f));
    BeamLight->SetIntensity(5000.0f);
    BeamLight->SetLightColor(FLinearColor(0.0f, 0.0f, 1.0f));
    BeamLight->SetVisibility(false);

    Speed = 100.0f; // Default speed
    SpeedVariance = 20.0f; // Default speed variance
    DirectionVariance = 15.0f; // Default direction variance in degrees
    DescentFrequency = 2.0f; // Default descent frequency in seconds
    DescentSpeed = 50.0f; // Default descent speed
    DescentDistance = 100.0f; // Default descent distance
    AltitudeChangeFrequency = 5.0f; // Default altitude change frequency in seconds
    MinAltitude = 300.0f; // Minimum altitude
    MaxAltitude = 700.0f; // Maximum altitude
    AbductionRange = 500.0f; // Default abduction range
    bIsFlying = false;
    bIsDescending = false;
    bIsAbducting = false;
    AbductingAnimal = nullptr;
}

// Called when the game starts or when spawned
void ASpaceshipActor::BeginPlay()
{
    Super::BeginPlay();
    
    // Set initial flight path
    FVector InitialStartLocation = FVector(100.0f, 100.0f, 500.0f);
    FVector InitialTargetLocation = FVector(1000.0f, 1000.0f, 500.0f);
    SetInitialFlightPath(InitialStartLocation, InitialTargetLocation);

    // Start descent process
    StartDescent();

    // Start random movement process
    StartRandomMovement();

    // Start altitude change process
    GetWorldTimerManager().SetTimer(AltitudeChangeTimerHandle, this, &ASpaceshipActor::ChangeAltitude, AltitudeChangeFrequency, true);

    // Start detection process
    GetWorldTimerManager().SetTimer(DetectionTimerHandle, this, &ASpaceshipActor::OnDetectAnimals, 1.0f, true);
}

// Called every frame
void ASpaceshipActor::Tick(float DeltaTime)
{
    Super::Tick(DeltaTime);

    if (bIsFlying)
    {
        FVector CurrentLocation = GetActorLocation();
        FVector NewLocation = CurrentLocation + (Direction * Speed * DeltaTime);
        
        // Check if the spaceship has reached or passed the target location
        if (FVector::Dist(NewLocation, TargetLocation) <= Speed * DeltaTime)
        {
            NewLocation = TargetLocation;
            bIsFlying = false;
            UE_LOG(LogTemp, Log, TEXT("Spaceship has reached the target location."));
        }

        SetActorLocation(NewLocation);
    }

    if (bIsDescending)
    {
        FVector CurrentLocation = GetActorLocation();
        FVector NewLocation = CurrentLocation + FVector(0.0f, 0.0f, -DescentSpeed * DeltaTime);

        // Check if the spaceship has reached or passed the descent target location
        if (CurrentLocation.Z - NewLocation.Z >= DescentDistance)
        {
            NewLocation.Z = CurrentLocation.Z - DescentDistance;
            bIsDescending = false;
            UE_LOG(LogTemp, Log, TEXT("Spaceship has completed a descent."));
        }

        SetActorLocation(NewLocation);
    }

    if (bIsAbducting && AbductingAnimal)
    {
        AnimateAbduction(DeltaTime);
    }
}

void ASpaceshipActor::SetInitialFlightPath(FVector StartLocation, FVector InTargetLocation)
{
    SetActorLocation(StartLocation);
    TargetLocation = InTargetLocation;
    Direction = (TargetLocation - StartLocation).GetSafeNormal();
    bIsFlying = true;
}

void ASpaceshipActor::StartDescent()
{
    bIsDescending = true;
    DescentTargetLocation = GetActorLocation() + FVector(0.0f, 0.0f, -DescentDistance);
}

void ASpaceshipActor::StartRandomMovement()
{
    GetWorldTimerManager().SetTimer(FlightPatternTimerHandle, this, &ASpaceshipActor::ChangeFlightPattern, DescentFrequency, true);
}

void ASpaceshipActor::ChangeFlightPattern()
{
    // Randomize direction slightly
    FVector NewDirection = Direction + FMath::VRand() * DirectionVariance;
    Direction = NewDirection.GetSafeNormal();
}

void ASpaceshipActor::ChangeAltitude()
{
    float NewAltitude = FMath::RandRange(MinAltitude, MaxAltitude);
    FVector NewLocation = GetActorLocation();
    NewLocation.Z = NewAltitude;
    SetActorLocation(NewLocation);
}

void ASpaceshipActor::OnDetectAnimals()
{
    DetectAnimals();

    if (DetectedAnimals.Num() > 0)
    {
        StartAbductionProcess(DetectedAnimals[0]);
    }
}

void ASpaceshipActor::DetectAnimals()
{
    FVector Location = GetActorLocation();
    TArray<FOverlapResult> OverlapResults;
    FCollisionShape Sphere = FCollisionShape::MakeSphere(AbductionRange);
    
    bool bHasOverlaps = GetWorld()->OverlapMultiByChannel(
        OverlapResults,
        Location,
        FQuat::Identity,
        ECC_Pawn,
        Sphere
    );

    if (bHasOverlaps)
    {
        DetectedAnimals.Empty();

        for (auto& Result : OverlapResults)
        {
            AActor* OverlappedActor = Result.GetActor();
            if (OverlappedActor)
            {
                DetectedAnimals.Add(OverlappedActor);
                UE_LOG(LogTemp, Log, TEXT("Detected animal: %s"), *OverlappedActor->GetName());
            }
        }
    }

    // Debug visualization
    DrawDebugSphere(GetWorld(), Location, AbductionRange, 50, FColor::Green, false, 1.0f);
}

void ASpaceshipActor::StartAbductionProcess(AActor* Animal)
{
    if (!bIsAbducting)
    {
        AbductingAnimal = Animal;
        bIsAbducting = true;

        // Activate abduction beam and light
        AbductionBeam->SetVisibility(true);
        BeamLight->SetVisibility(true);

        UE_LOG(LogTemp, Log, TEXT("Starting abduction process for animal: %s"), *Animal->GetName());
    }
}

void ASpaceshipActor::AnimateAbduction(float DeltaTime)
{
    if (AbductingAnimal)
    {
        FVector AnimalLocation = AbductingAnimal->GetActorLocation();
        FVector SpaceshipLocation = GetActorLocation();

        FVector DirectionToSpaceship = (SpaceshipLocation - AnimalLocation).GetSafeNormal();
        float AbductionSpeed = 100.0f;

        FVector NewLocation = AnimalLocation + (DirectionToSpaceship * AbductionSpeed * DeltaTime);
        AbductingAnimal->SetActorLocation(NewLocation);

        // Check if the animal has reached the spaceship
        if (FVector::Dist(NewLocation, SpaceshipLocation) <= AbductionSpeed * DeltaTime)
        {
            bIsAbducting = false;

            // Deactivate abduction beam and light
            AbductionBeam->SetVisibility(false);
            BeamLight->SetVisibility(false);

            // Remove the animal from the scene
            RemoveAbductedAnimal();
        }
    }
}

void ASpaceshipActor::RemoveAbductedAnimal()
{
    if (AbductingAnimal)
    {
        UE_LOG(LogTemp, Log, TEXT("Removing abducted animal: %s"), *AbductingAnimal->GetName());

        // Destroy the animal actor
        AbductingAnimal->Destroy();

        // Remove the animal from the detected animals list
        DetectedAnimals.Remove(AbductingAnimal);

        // Clear the abduction reference
        AbductingAnimal = nullptr;
    }
}
```
FILENAME: AnimalActor.h
```cpp
#pragma once

#include "CoreMinimal.h"
#include "GameFramework/Actor.h"
#include "AnimalActor.generated.h"

UCLASS()
class A00_API AAnimalActor : public AActor
{
    GENERATED_BODY()

public:
    AAnimalActor();

protected:
    virtual void BeginPlay() override;

public:
    virtual void Tick(float DeltaTime) override;

private:
    UPROPERTY(VisibleAnywhere)
    UStaticMeshComponent* AnimalMesh;
};
```
FILENAME: AnimalActor.cpp
```cpp
#include "AnimalActor.h"
#include "Components/StaticMeshComponent.h"
#include "UObject/ConstructorHelpers.h"

AAnimalActor::AAnimalActor()
{
    PrimaryActorTick.bCanEverTick = true;

    AnimalMesh = CreateDefaultSubobject<UStaticMeshComponent>(TEXT("AnimalMesh"));
    static ConstructorHelpers::FObjectFinder<UStaticMesh> AnimalMeshAsset(TEXT("/Game/StarterContent/Shapes/Shape_Cube.Shape_Cube"));
    if (AnimalMeshAsset.Succeeded())
    {
        AnimalMesh->SetStaticMesh(AnimalMeshAsset.Object);
        RootComponent = AnimalMesh;
        AnimalMesh->SetRelativeScale3D(FVector(0.5f, 0.5f, 0.5f)); // Scale down the cube to represent an animal
    }
    else
    {
        UE_LOG(LogTemp, Warning, TEXT("Animal mesh not found!"));
    }
}

void AAnimalActor::BeginPlay()
{
    Super::BeginPlay();
}

void AAnimalActor::Tick(float DeltaTime)
{
    Super::Tick(DeltaTime);
}
```
Here is the layout of the scene:

CLASS: FarmEnvironmentActor
LOCATION: (0.0, 0.0, 0.0)
ROTATION: (0.0, 0.0, 0.0)

CLASS: SpaceshipActor
LOCATION: (100.0, 100.0, 500.0)
ROTATION: (0.0, 0.0, 0.0)

CLASS: AnimalActor
LOCATION: (150.0, 150.0, 0.0)
ROTATION: (0.0, 0.0, 0.0)

CLASS: AnimalActor
LOCATION: (200.0, 200.0, 0.0)
ROTATION: (0.0, 0.0, 0.0)

CLASS: AnimalActor
LOCATION: (300.0, 300.0, 0.0)
ROTATION: (0.0, 0.0, 0.0)

CLASS: AnimalActor
LOCATION: (600.0, 600.0, 0.0)
ROTATION: (0.0, 0.0, 0.0)

CLASS: AnimalActor
LOCATION: (1000.0, 1000.0, 0.0)
ROTATION: (0.0, 0.0, 0.0)
\end{lstlisting}

\section{Prompt generation code}

Here we provide the code snippets responsible for generating system prompts for \textit{DreamGarden}'s various modules.

\subsection{Planner}

Here is the python code used to generate system prompts for the planner module (both when transforming the initial user seed prompt to a broad plan, and when recursively breaking down the plan into more fine-grained steps). Note that the maximum branching factor and depth are specified in the config `cfg`. In our experiments, `cfg.test\_disclaimer` is always `True`.

\pythonexternal{prompts/planner_system_prompts.py}
\label{lst:planner_prompts}

Here is the code used to generate prompts based on user input and the existing plan tree.

\pythonexternal{prompts/planner_prompts.py}

\subsection{Task generator}

The following code produces prompts used by the task generator, which is responsible for transforming leaf nodes in the tree generated by the planner into concrete implementation tasks (i.e. into prompts for specific implementation submodules). Note that this code makes use of the `available\_submodules` dictionary defined above.

\pythonexternal{prompts/task_generator.py}

\subsection{Code generator}

The following code produces prompts for generating/repairing/extending code, including in cases where the code should involve a procedural mesh.

\pythonexternal{prompts/code_generator.py}

The following code produces prompts for evaluating the compilation logs produced when compiling generated code, and providinv feedback to the code generator in case of compilation failure.

\pythonexternal{prompts/code_compile_eval.py}

The following code produces prompts for evaluating the Unreal Engine crash logs---in case a run-time crash occurs when running generated code---and providing feedback to the code generator.

\pythonexternal{prompts/crash_eval.py}

The following code produces prompts for evaluating visual output and runtime logs in case generated code compiles and runs without error.

\pythonexternal{prompts/viz_eval.py}

\newpage
\section{User Study}

\subsection{Intake survey}

\begin{itemize}
\item What is your level of familiarity with Generative AI systems like ChatGPT or DALL-E Image Generator?
\item Can you briefly describe your experience using Generative AI systems? (N/A if not applicable)
\item What is your level of familiarity with game design and/or game engines?
\item Can you briefly describe your experience with game design and/or game engines?
\item (N/A if not applicable)	What is your level of familiarity with Unreal Engine?
\item Can you briefly describe your experience with Unreal Engine? (N/A if not applicable)
 \end{itemize}
 
Here, questions asking for a user's level of familiarity have multiple choice answers with the following options:
\begin{itemize}
    \item Very unfamiliar
    \item Unfamiliar
    \item Slightly unfamiliar
    \item Slightly familiar
    \item Familiar
    \item Very familiar
\end{itemize}

Other questions allowed for short responses.

\subsection{Exit survey}

\begin{itemize}
\item In what scenarios would you find it valuable to use a tool like LucidDev\footnote{Note that during the user study, the system had the working name ``LucidDev'', which was later changed to ``DreamGarden''.}?
\item Are there any situations where you would prefer not to use a tool like LucidDev? If so, why?
\item What aspects of the system, if any, did you find appealing/enjoyable? What aspects did you find unappealing/unenjoyable?	
\item Reflecting on LucidDev in its current state, imagine a more fully-fledged version of this system. In a perfect world, what would this system look like? What features would the system have that it currently lacks?
\item Brainstorm positive and negative implications of your vision. Describe them here.
\item On a scale from 1-5 (5 being most enjoyable) how enjoyable did you find using the tool?
\item On a scale from 1-5 (5 being most frustrating) how frustrating did you find using the tool?
\end{itemize}

All questions in the exit survey allowed for short responses from users.

\end{document}